\documentstyle[12pt,titlepage]{article}

\renewcommand{\theequation}{\thesection.\arabic{equation}}
\def\bl{\Biggl\{}
\def\br{\Biggr\}}
\def\bpl{\Biggl(}
\def\bpr{\Biggr)}
\def\bbl{\Biggl[}
\def\bbr{\Biggr]}
\def\bb{\Biggl|}

\def\fl{\flushleft}
\def\L{{\cal L}}

\def\a{\alpha}

\def\d{\delta}
\def\D{\Delta}
\def\e{\epsilon}

\def\M{{\cal M}}
\def\O{{\cal O}}
\def\g{\gamma}
\def\gbig {\hbox{\Large\it g}}
\def\J{{\cal J}}
\def\k{\kappa}
\def\l{\lambda}

\def\so{\emptyset}

\def\z{\zeta}

\def\s{\sigma}

\def\z{\zeta}

\def\vphi{\varphi}
\def\w{\omega}

\def\hf{\frac{1}{2}}
\def\der{\partial}

\def\bq{\begin{equation}}
\def\eq{\end{equation}}
\def\brr{\begin{eqnarray}}
\def\err{\end{eqnarray}}
\def\ba{\left(\begin{array}}
\def\ea{\end{array}\right)}

\def\lapp{\stackrel{<}{\sim}}

\def\ba{\left(\begin{array}}
\def\ea{\end{array}\right)}
\input epsf.tex
\epsfverbosetrue
\begin{document}
\pagestyle{empty}
\begin{flushright} CERN-TH.7301/94, UPR-608T \end{flushright}
\begin{large}
\begin{center}{\bf INSTANTON EFFECTS IN MATRIX MODELS}\end{center}
\begin{center}{\bf AND STRING EFFECTIVE LAGRANGIANS}\end{center}

\end{large}
\begin{center} RAM BRUSTEIN  \end{center}
\vspace{-.3in}
\begin{center} Theory Division, CERN \end{center}
\vspace{-.3in}
\begin{center} CH-1211 Geneva 23, Switzerland  \end{center}

\begin{center} MICHAEL FAUX$^{*)}$ and BURT A. OVRUT$^{*)}$
\end{center}
\vspace{-.3in}
\begin{center} Department of Physics \end{center}
\vspace{-.3in}
\begin{center} University of Pennsylvania \end{center}
\vspace{-.3in}
\begin{center} Philadelphia, Pa 19104-6396 \end{center}
\begin{center} ABSTRACT \end{center}

We perform an explicit  calculation of the lowest order effects of
single eigenvalue instantons on the continuous sector of the collective
field
theory derived from a $d=1$ bosonic matrix model. These effects consist
of
certain induced operators whose exact form we exhibit.

\vspace*{1.8cm}
\noindent
\rule[.1in]{16.5cm}{.002in}\\
\noindent
$^{*)}$ Work supported in part by DOE under Contract No.
DOE-AC02-76-ERO-3071.
\vspace*{0.5cm}

\begin{flushleft} CERN-TH.7301/94 \\
June 1994
\end{flushleft}
\vfill\eject

\setcounter{page}{1}
\pagestyle{plain}

\renewcommand{\theequation}{1.\arabic{equation}}
\setcounter{equation}{0}
{\fl{\bf 1. Introduction}}

Recently, it has been shown that matrix models \cite{mm} allow the
construction of space-time Lagrangians valid to all orders in the
string
coupling parameter,
at least for noncritical strings propagating in $d=2$ dimensions.
These Lagrangians are derived using the techniques of collective field
theory
\cite{collft,jevrev}.   All order  Lagrangians have been constructed,
using
these techniques, for both the $d=1$ bosonic matrix model \cite{bo} and
also
for the $d=1, {\cal N}=2$ supersymmetric matrix model \cite{bfo}.
There are two remarkable features of these constructions.  First, when
interactions are included to all orders, the
induced coupling blows up at finite points in space and delineates a
zone of
strong coupling.  This is to be contrasted with the lowest order
theory, where
the coupling only diverges at spatial infinity.  Secondly, since these
all-order Lagrangians are derived from matrix models, they contain
additional
non-perturbative information which is directly
accessible and computable.   The existence of these new
non-perturbative
aspects of the theory relies on the observation
that the matrix models contain two distinct sectors.  The first of
these is the so-called continuous sector, which consists of a
continuous
distribution of
matrix eigenvalues. The second sector consists of discrete eigenvalues,
which
are
distinguishable from the continuum eigenvalues.  The classical
configurations
of the matrix
model include time-dependent instanton solutions in which the discrete
eigenvalues tunnel between two continuous eigenvalue sectors. In this
paper we
perform an explicit  calculation of the leading order effects of such
single
eigenvalue instantons on the effective theory derived
from a $d=1$ bosonic matrix model.
These consists of a set of induced operators, whose exact form we
compute and
exhibit.

This work is particularly relevant for the following
reason.  It is conjectured that, in the supersymmetric case, the same
instantons described in this paper, and their
associated bosonic and fermionic zero modes, provide a mechanism
for supersymmetry breaking in the associated $d=2$ effective
superstring
theory. It is presumed that the discrete nature of the
single eigenvalues allows a novel circumvention of a particular no-go
theorem,
based on Witten's index, relevant to non-perturbative supersymmetry
breaking in
$d>1$ dimensions.  The calculation in this paper is  a necessary
preliminary to
the explicit calculation of this effect,
which will be discussed in a forthcoming paper \cite{nbfo}.
Non-perturbative effects due to single eigenvalue instantons
were also discussed elswhere \cite{shenker,dabh,leemend}

This paper is structured as follows.

In section 2 we describe the
distinct sectors of the matrix model in some detail.  We compute
the equation of motion for both the continuous sector and for the
discrete sector and we  analyse the mutual interaction
between these two.  We then compute and
exhibit the complete set of single eigenvalue instanton
solutions valid to lowest order in a  small coupling constant.

In section 3, we integrate out  the single eigenvalue
instantons in a dilute-gas approximation.  This then gives
rise to a collective  field theory which has the instanton effects
incorporated.

\renewcommand{\theequation}{2.\arabic{equation}}
\setcounter{equation}{0}
{\fl{\bf 2. Bosonic Matrix Models}}

A $d=1$ bosonic matrix model has a time-dependent $N\times N$ Hermitian
matrix, $M(t)$, as its fundamental variable.  Its dynamics are
described by the
Lagrangian
\bq L(\dot{M},M)=\hf Tr \dot{M}^2-V(M).
 \label{lagm}\eq
The potential is taken to be polynomial,
\bq V(M)=\sum_{n=0}^\infty a_n Tr M^n,
 \label{v1def}\eq
where the $a_n$ are real coupling parameters.
The mass dimension of $M$ is $\hf$ and the $a_n$ have
positive mass dimension $(n+2)/2$. As $N\rightarrow\infty$, if the
$a_n$ are
tuned simultaneously and
appropriately, the associated partition function describes an
ensemble of oriented two-dimensional
Riemann surfaces, including contributions at all genus.
It is argued that, in this limit,
the model describes a string propagating in two
space-time dimensions.
For this to be so, it is necessary that the $a_n$ scale as $N^{1-n/2}$
for large $N$.
We will, henceforth, assume that the coupling parameters scale in this
manner.  It follows that, in the large $N$ limit, all terms in
(\ref{v1def}) with $n\ge 3$ become negligibly small.  Furthermore,
the $n=1$ term can be shifted away and the remaining terms in the
potential written as
\bq V(M)=Tr(N{V}_0\cdot {\bf 1}-\hf\w^2M^2),
 \label{vdef}\eq
where ${\bf 1}$ is the
$N\times N$ unit matrix.  The parameters ${V}_0$ and $\w$ each
have mass dimension one, and are arbitrary.  In (\ref{vdef}) the
scaling
behavior of the coefficients has been made explicit.
The Lagrangian, (\ref{lagm}), is invariant under the
global $U(N)$ transformation $M\rightarrow{\cal U}^\dagger M{\cal U}$,
where ${\cal U}$ is an arbitrary $N\times N$ unitary matrix.  The set
of
states which do not transform under ${\cal U}$ comprise the
$U(N)$-singlet sector of the quantized theory.
It can be shown that the physics of this singlet sector is described
equivalently by a theory involving only the $N$ eigenvalues, $\l_i(t)$,
of the
matrix $M(t)$ with the following Lagrangian,
\bq L[\l]=\sum_{i=1}^N\{\hf\dot{\l}_i^2-({V}_0-\hf\w^2\l_i^2)
    -\hf\sum_{j\ne i}\frac{1}{(\l_i-\l_j)^2}\}.
 \label{lageig}\eq
The eigenvalues are first restricted to lie in the interval
$-\frac{L}{2}\le \l_i\le\frac{L}{2}$ for any $i$.
When we take the limit
$N\rightarrow\infty$, we will simultaneously take
$L\rightarrow\infty$.  In this limit, over a given range, $l$,
to be made explicit below,
there exist two possibilities.
If $n$ represents the number of eigenvalues within this range,
then the average density is given by $\rho=n/l$.  In the limit
$N\rightarrow\infty, L\rightarrow\infty$, $\rho$ can remain small,
and the eigenvalues populate the region sparsely.
We refer to this situation
as a ``low density" or ``discrete" distribution of eigenvalues over the
region
$l$.
In the second case, $\rho$ becomes very
large or infinite, and the eigenvalues
populate the region densely. In this case, the eigenvalues can
be aggregated into a ``collective field" which describes their physics
en-masse.  We refer to this second case as a ``high density"
or``continuous"
distribution
of eigenvalues.
We begin by studying the continuous case.\\

{\fl{\it 2.1 Collective Field Theory}}\\
\indent
As yet, $N$ and $L$ remain finite.
We introduce a continuous
real parameter, $x$, constrained to lie in the interval
$-\frac{L}{2}\le x\le\frac{L}{2}$, and over this line segment define a
collective
field,
\bq \der_x\vphi(x,t)=\sum_{i=1}^N\d(x-\l_i(t)).
 \label{pdef}\eq
Since the $\l_i$ have mass dimension $-\hf$, the parameter
$x$ also has mass dimension $-\hf$.  The delta function has the inverse
dimensionality of its argument, which is $+\hf$.  Thus, since $\der_x$
is also
a dimension $+\hf$ operator, it follows that $\vphi(x,t)$ is
dimensionless.
It follows from (\ref{pdef}) that
\bq \int_{x_0}^{x_0+l}dx\der_x\vphi(x,t)=n,
 \label{rel}\eq
where $n$ is the number of eigenvalues in the range $l$.
Thus, $\vphi'=\der_x\vphi$ is
the eigenvalue density.  In the range $l$,
$\vphi'$ has $n$ degrees of freedom.  Provided that
$n/l\rightarrow\infty$ as $N\rightarrow\infty, L\rightarrow\infty$,
the average density of eigenvalues then becomes infinite, and,
modulo some technical subtleties irrelevant to this discussion,
the field $\vphi$ becomes an unconstrained, ordinary
two dimensional field.
In effect,
$\vphi'$ ceases to be a sum over delta
functions and becomes a continuous eigenvalue density.
It can be shown, in this case,  that
the eigenvalue Lagrangian, (\ref{lageig}), may be rewritten in terms of
the
collective field as follows,
\bq L[\vphi]=\int
dx\{\frac{\dot{\vphi}^2}{2\vphi'}-\frac{\pi^2}{6}\vphi^{'3}
    -(V_0-\frac{\w^2}{2}x^2)\vphi'\}.
 \label{lagcoll}\eq
The associated action is given by $S[\vphi]=\int dt L[\vphi]$.
This expression describes the physics over all ranges of $x$ where the
eigenvalue density is large.  The limits on the $\int dx$ integral
are set accordingly.  In this subsection, we restrict our attention
to such a continuous
sector.  In the next subsection we will discuss the incorporation
of a discrete sector into the theory.
Since our interest is in the quantum theory, henceforth we will
consider only the Euclidean version of the action, which is given by
\bq S_E[\vphi]=\int dx dt \{\frac{\dot{\vphi}^2}{2\vphi'}
    +\frac{\pi^2}{6}\vphi^{'3}
    +(V_0-\frac{\w^2}{2}x^2)\vphi'\}.
 \label{se}\eq
The
equation of motion, obtained by varying
(\ref{se}) is
\bq \der_t(\frac{\dot{\vphi}}{\vphi'})
    -\hf\der_x\bl\frac{\dot{\vphi}^2}{\vphi^{'2}}+\pi^2\vphi^{'2}
    -\w^2x^2\br=0.
 \label{momo}\eq
The static solution is obtained by taking $\dot{\vphi}=0$, so that
(\ref{momo})
reduces to
\bq \der_x\bl\pi^2\vphi^{'2}-\w^2x^2\br=0.
 \label{eqmo}\eq
The most general solution to this equation is the following,
\bq \tilde{\vphi}_0^{'}(x)=\frac{\w}{\pi}\sqrt{x^2-A^2},
 \label{vback}\eq
where $A^2$ is a constant which can be negative, zero, or positive.
Since (\ref{se}) only involves derivative couplings, however,
the equation of motion, (\ref{eqmo}),
is not sufficient to extremize the action. This is because
the action depends on the value of $A^2$, which is undetermined
by (\ref{eqmo}).
In order to determine $A^2$ we must compute the action using
(\ref{vback}) and find the value which represents the true
extremum.
Inserting (\ref{vback}) into (\ref{se}),
we find, for finite $L$, over a finite Euclidean time duration,
$-\frac{T}{2}\le t\le\frac{T}{2}$, that
\brr S_E[\tilde{\vphi}_0]
     &=& -\frac{\w^3T}{96\pi}
     \bl (L^2+2A^2-24\frac{V_0}{\w^2})L\sqrt{L^2-4A^2} \nonumber \\
     & & \hspace{.5in}
     -24A^2(A^2-4\frac{V_0}{\w^2})\ln
     (\frac{L+\sqrt{L^2-4A^2}}{2|A|})\br.
 \label{sep}\err
This function is minimized, for all values
of $L, T, V_0$, and $\w$ when
\bq A^2=2V_0/\w^2.
 \label{aak}\eq
This determines $A^2$ in terms of the two parameters, $V_0$ and $\w$,
but
its sign remains undetermined. For convenience we will write $V_0$ as
$\hf\w^2A^2$.
In Figure 1, we plot (\ref{vback}) for the three cases
$A^2<0, A^2=0$, and $A^2>0$.
Superimposed on this plot is the
``potential", $V=\hf\w^2(A^2-x^2)$ which multiplies the linear $\vphi'$
term
in (\ref{lagcoll}).

\ \vspace{1pc}

\epsfxsize=480pt
\centerline{\epsfbox{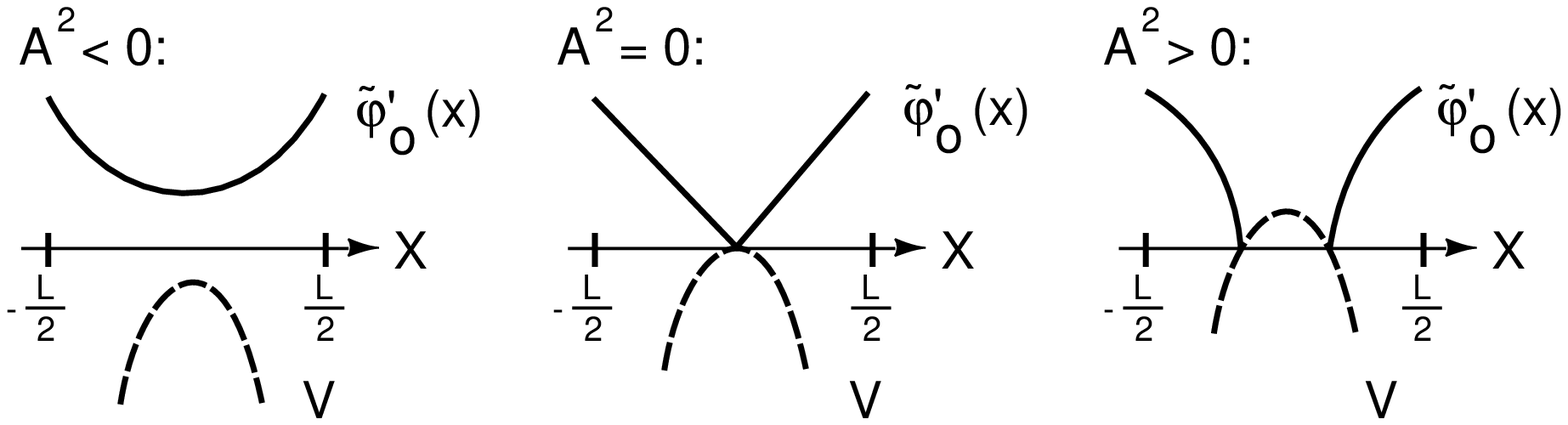}}
\centerline{\tenrm Figure~1. The potential and classical  solution for
different values of $A^2$.}
\noindent
For the cases $A^2\le 0$, the
eigenvalues continuously populate all values.
That is, the range $l$ over which there is a continuous distribution
of eigenvalues is given by $-\frac{L}{2}\le x\le \frac{L}{2}$.
The case $A^2>0$, however,
leaves a region, $|x|<A$, which is not continuously
occupied by eigenvalues, where a discrete sector may be
accomodated.
In this case the range $l$ over which there is a continuous
distribution
of eigenvalues is given by $-\frac{L}{2}\le x\le -A$ and $A\le x \le
\frac{L}{2}$.
Our interest in this paper is to develop a technique
for systematically encorporating discrete eigenvalue dynamics into the
collective field theory.  We therefore restrict attention, for the
remainder of this paper, to the case $A^2>0$.

Since $\vphi'$ is now a continuous density of eigenvalues, we may use
(\ref{rel}) to determine the approximate location of the
first eigenvalues in the continuum; that is, those
two eigenvalues closest to $x=\pm A$.
We focus on the region $x\ge A$.
There is an identical discussion regarding the opposite region,
$x\le-A$.
Given (\ref{vback}), the first eigenvalue must live
somewhere in the region $A\le x\le A+\e_x$, where $\e_x$ is determined
by
the following relation,
\brr 1 &=& \frac{\w}{\pi}\int_A^{A+\e_x}dx\sqrt{x^2-A^2} \nonumber \\
       &=& \frac{\w A^2}{2\pi}\bl\frac{x}{A}\sqrt{(\frac{x}{A})^2-1}-
       \ln(\frac{x}{A}+\sqrt{(\frac{x}{A})^2-1})\br
       \Biggl{|}_{x=A}^{x=A+\e_x}.
 \label{ky}\err
We make the important assumption that $\e_x<<A$.  After some algebra,
Eq.(\ref{ky}) then becomes
\bq \hf(\frac{3\pi}{\w A^2})^{2/3}=\frac{\e_x}{A}
    +{\cal O}\bpl(\frac{\e_x}{A})^2\bpr.
 \eq
For consistency, this requires that $(\w A^2)^{-1}<<1$.
This small dimensionless number will be central to much of the ensuing
analysis, so we give it a special name,
\bq g=\frac{1}{\w A^2}<<1.
 \label{gdef}\eq
Since $\tilde{\vphi}_0^{'}$
increases monotonically
as $x$ becomes larger than $A$, it is reasonable to assume that the
first eigenvalue actually has
a value nearer to $x=A+\e_x$ rather than nearer to $x=A$.
At any rate, it is clear that the first eigenvalue does not live
precisely at the value $x=A$.  This distinction will prove a necessary
and important regulator on quantities which we will encounter.  For
definiteness, we assume henceforth that the first eigenvalue in the
static
continuum has a value $x=A+\e_x$, where
\bq \e_x=\hf(3\pi g)^{2/3}A
 \label{edef}\eq
and $g$ is a small, dimensionless number, which,
in the present context, parameterizes the width of the discrete region
as well
as our ignorance
regarding the ``graininess" of eigenvalues near the edge of the
continuous
distribution, when we adopt a collective
field point of view.

{\fl{\it 2.2 Discrete Eigenvalue Dynamics}}\\
\indent
We now turn our attention to the region $|x|\le A$.
We assume, in addition to a continuum of eigenvalues
$\l_i$ for $i=1$ to $N$, that there exists an additional
discrete eigenvalue, which we
denote $\l_0$.   There are then $N+1$
total eigenvalues, and the Euclidean version of Lagrangian
(\ref{lageig}) now reads
\brr L_E = \sum_{i=0}^N\{\hf\dot{\l}_i^2+({V}_0-\hf\w^2\l_i^2)
    +\hf\sum_{j\ne i}\frac{1}{(\l_i-\l_j)^2}\}.
 \err
Note that the index $i$ now runs over the $N+1$ values from $0$ to $N$.
What do we mean by a discrete eigenvalue?  It was shown in the previous
section that the separation of the continuum eigenvalues nearest to
$\pm A$ is of order $\e_x$.  As long as $-A\le\l_0\le A$, and
\bq A-|\l_0|>>\e_x,
 \label{condo}\eq
the eigenvalue $\l_0$ is truly distinct from the continuum and,
hence, discrete.
Assuming that $\l_0$ satisfies (\ref{condo}),
it is useful to rewrite this Lagrangian by separating the $\l_0$
contribution
from the contribution due to the continuum eigenvalues, as follows,
\brr L_E &=& \hf\dot{\l}_0^2+(V_0-\hf\w^2\l_0^2)
     +\sum_{i\ne 0}\frac{1}{(\l_0-\l_i)^2} \nonumber \\
     &+&\sum_{i=1}^N\{\hf\dot{\l}_i^2+({V}_0-\hf\w^2\l_i^2)
     +\hf\sum_{j\ne i}\frac{1}{(\l_i-\l_j)^2}\}.
\err
As above, we may now rewrite this expression using the definition
(\ref{pdef}).  We thus obtain
\brr L_E[\l_0;\vphi] &=& \hf\dot{\l}_0^2+\hf\w^2(A^2-\l_0^2)
     +\int dx\frac{\vphi'}{(x-\l_0)^2} \nonumber \\
     &+& \int dx\{\frac{\dot{\vphi}^2}{2\vphi'}
     +\frac{\pi^2}{6}\vphi^{'3}
     +\hf\w^2(A^2-x^2)\vphi'\}.
 \label{hybrid}\err
The third term in this expression represents the mutual
interaction of the discrete eigenvalue with the continuum eigenvalues,
which are collectively described using the field $\vphi$.
We obtain the Euclidean equations of motion for $\l_0$ and for $\vphi$
by variation of (\ref{hybrid}).  Respectively, these are found to be
\bq \ddot{\l}_0+\w^2\l_0+\int dx\frac{\vphi'}{(\l_0-x)^3}=0
 \label{mo1}\eq
\bq \der_t(\frac{\dot{\vphi}}{\vphi'})
    -\hf\der_x\bl\frac{\dot{\vphi}^2}{\vphi^{'2}}+\pi^2\vphi^{'2}
    -\w^2x^2+\frac{2}{(\l_0-x)^2}\br=0.
 \label{mo2}\eq
We consider first the $\vphi$ equation.
We proceed to show, even in the presence of a nontrivial,
but discrete, $\l_0(t)$,
that the static background, $\tilde{\vphi}_0^{'}$, derived above
is still a valid solution to leading order in $\e_x$.
In order that $\tilde{\vphi}_0^{'}$ remains a valid
solution, it must be so that the last term on the left hand
side of (\ref{mo2}) is negligible with respect to the two which precede
it.  We can then consistently neglect the time-dependent part of
(\ref{mo2})
as well.  Since $\pi^2\tilde{\vphi}_0^{'}-\w^2x^2=-\w^2A^2$,  this
requirement is that
\bq \w^2A^2>>(\l_0-x)^{-2}.
 \label{cond}\eq
Furthermore, since $\l_0$ satisfies (\ref{condo}),
$\e_x^{-2}\ge (\l_0-x)^{-2}$ and
from (\ref{gdef}) and (\ref{edef}) we derive
$\e_x^{-2}\approx\w^2A^2\cdot g^{1/3}$.  Therefore, condition
(\ref{cond}) requires simply that $\w^2A^2>>\e_x^{-2}$ or, equivalently
\bq g^{1/3}<<1,
 \eq
which we have already assumed.
With this discussion in mind, we regard (\ref{vback}) as the static
solution
to (\ref{mo2}), despite the presence of
an additional discrete eigenvalue.
We discuss below exacly how it is that such a discrete
eigenvalue can arise.

Next, we turn our attention to the $\l_0$ equation, (\ref{mo1}).  This
is
the Euclidean equation of motion,
\bq \ddot{\l}_0-V_{eff}^{'}(\l_0)=0,
 \label{moe}\eq
where
\bq V_{eff}(\l_0)=-\hf\w^2\l_0^2+\widehat{V}(\l_0).
 \label{vef}\eq
In this expression, $\widehat{V}$ is the mean field interaction of
$\l_0$
with all of the continuum eigenvalues,
\bq \widehat{V}(\l_0)=(\int_{-L/2}^{-A}dx+\int_A^{L/2}dx)
    \frac{\tilde{\vphi}_0^{'}}
    {(\l_0-x)^2}.
 \eq
Using (\ref{vback}) and (\ref{aak}), we can compute this function for
finite $L$.  Ignoring an irrelevant constant term, and using
(\ref{gdef}),
the full effective potential, in the limit $L\rightarrow\infty$ is
then found to be
\bq V_{eff}(\l_0)=\frac{\w}{2g}\bl
    -(\frac{\l_0}{A})^2+4g\frac{(\l_0/A)}{\sqrt{1-(\l_0/A)^2}}
    \tan^{-1}(\frac{(\l_0/A)}{\sqrt{1-(\l_0/A)^2}})\br.
 \label{vvv}\eq
This function is plotted in Figure 2 for three different values of
$g$.
\ \vspace{1pc}

\epsfxsize=300pt
\centerline{\epsfbox{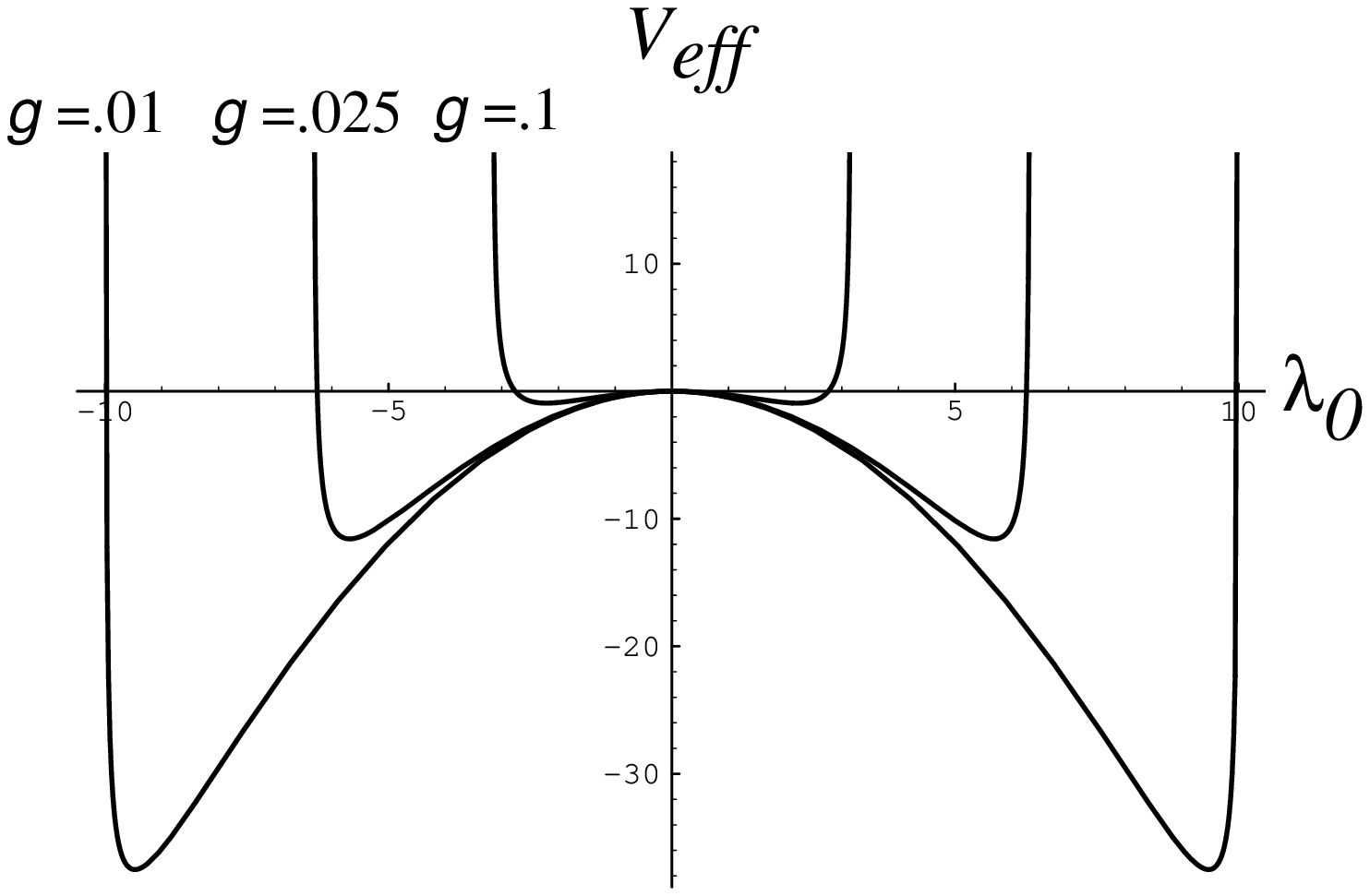}}
\centerline{\tenrm Figure~2. Effective Potential for
$g=.1,.025$ and $.01$. $V_{eff}$ and $\l_0$ are in units in which
$\w=1$.}
\noindent
It is clear from the figure that the effect of the second term in
(\ref{vvv}),
is to
turn the potential over near $\l_0=\pm A$, where it adds infinite
confining walls.  For small values of $g$,
 the minima of (\ref{vvv}) occur at
$\l_0=\pm(A-3\sqrt{3}\e_x)$ to leading order in $\e_x$.
However, we must be careful.
Recall that $\l_0$ is a discrete eigenvalue,
and $V_{eff}(\l_0)$ is well defined, only if $\l_0$ satisfies the
condition
(\ref{condo}).  It is clear that these minima do not satisfy this
condition
and, hence lie outside the range of validity of our
approximation.
The actual situation is the following.  As we have said, eigenvalue
$\l_0$ is
discrete and separated from the continuum, and $V_{eff}(\l_0)$ is well
defined,
provided $\l_0$ satisfies (\ref{cond}); that is, if $\l_0$ is
sufficiently far
from $\pm A$.  However, when $\l_0$ approaches $\pm A$ to within order
$\e_x$ it, in effect, enters the continuum.  This is because its
separation
from the first eigenvalues of the continuum is of the same order as the
``graininess" of the continuum discussed previously.  Under these
circumstances,
all eigenvalues, including $\l_0$, must be treated as a continuum using
a single collective field with action (\ref{se}).  It follows that
there is only one equation of motion, the $\vphi$ equation given in
(\ref{momo}), whose static solution is shown in (\ref{vback}).  Thus,
the
true equilibrium positions for $\l_0$ are at $\l_0=\pm(A+\e_x)$ rather
than
at $\l_0=\pm(A-3\sqrt{3}\e_x)$
given above.  To conclude, $\l_0$ can be treated as
discrete, and $V_{eff}(\l_0)$ is well defined, for $\l_0$ sufficiently
far
from $\pm A$.  When $\l_0$ approaches $\pm A$ to within order $\e_x$
it is absorbed into the continuum, and disappears as a discrete entity.
Of
course, this process can be reversed.  It is possible for the first
eigenvalue
of the continuum to ``leak" out and become a discrete
eigenvalue $\l_0$.  We will return to such processes below.

This being said, we would like to find both static and time-dependent
solutions for the Euclidean $\l_0$ equation of motion (\ref{moe}).  As
will
become clear in the next section, we need only do this to lowest order;
that is, to order $\e_x^0$.  In this case, we may take $\l_0$ as
discrete, and
$V_{eff}(\l_0)$ as well defined, for all values of $\l_0$ in the range
$-A\le\l_0\le A$.  Of course, $V_{eff}(\l_0)$ must now be evaluated in
the
limit that $g\rightarrow 0$.  This limiting case is given by
$V_{eff}(\l_0)=-\hf\w^2\l_0^2$ for $-A<\l_0<A$.
At $\l_0=\pm A$, though, the potential turns over abruptly and
becomes infinite confining walls, as discussed above.
As $g\rightarrow 0$ the minima of the potential occur at
$\l_0=\pm A$, where the potential obtains cusps, which do not have
well defined derivatives.  For any finite value of $g$, however, the
derivative vanishes at the minima of the potential.  It is appropriate
then,
in the $g\rightarrow 0$ limit, to take
$V_{eff}^{'}(\pm A)=0$. Hence, in this limit we can
replace (\ref{moe}) by
\brr \ddot{\l}_0+\w^2\l_0 &=& 0
     \hspace{.3in}; \hspace{.1in} -A<\l_0<A \nonumber \\
     \ddot{\l}_0 &=& 0
     \hspace{.3in}; \hspace{.1in} \l_0=\pm A.
 \label{keq} \err
We also impose the following boundary conditions,
$\l_0(t\rightarrow -\infty)= \pm A$ and, independently,
$\l_0(t\rightarrow +\infty)= \pm A$.
There are two static solutions to (\ref{keq}) which satisfy this
boundary
condition,
\bq {\widehat\l}_{0\pm}=\pm A.
 \eq
A simple time-dependent solution is given
by
\bq {\widehat\l}_0^{(+)}(t;t_1)=\left\{
   \begin{array}{ll}
    -A & ;\hspace{.2in} t<t_1-\frac{\pi}{2\w} \\
    +A\sin\w(t-t_1) & ;\hspace{.2in}
    t_1-\frac{\pi}{2\w}\le t\le t_1+\frac{\pi}{2\w} \\
    +A & ;\hspace{.2in} t>t_1+\frac{\pi}{2\w} \end{array}\right.,
 \label{lop1}\eq
where $t_1$ is arbitrary.
The solution (\ref{lop1})
describes an eigenvalue which rolls (tunnels) from $-A$ to $+A$ over
a time interval of duration $\frac{\pi}{\w}$, centered at an arbitrary
time $t_1$.
This solution is shown picturially in Figure 3.
\ \vspace{1pc}

\epsfxsize=250pt
\centerline{\epsfbox{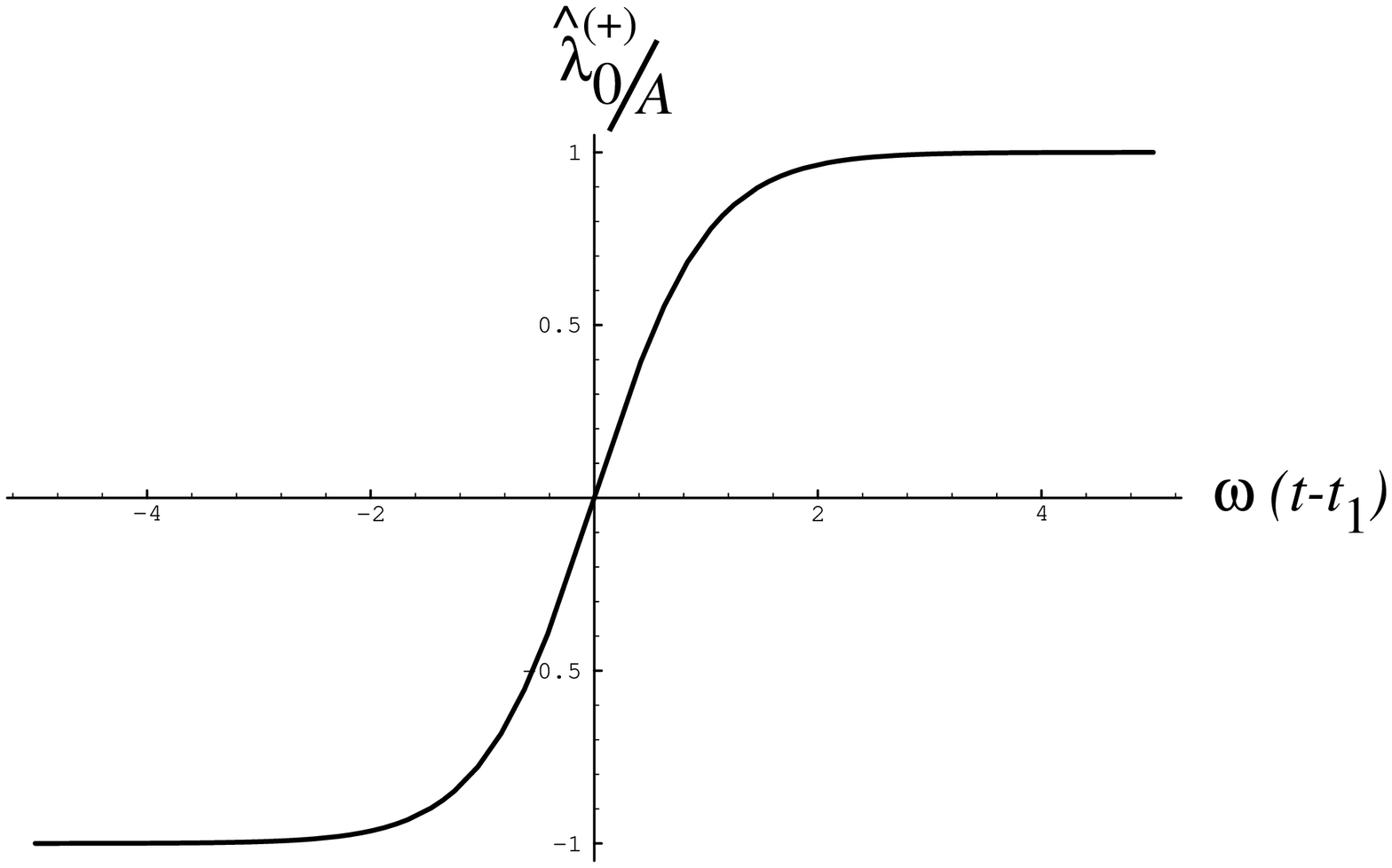}}
\centerline{\tenrm Figure~3. The ``kink" solution,
${\widehat\l}_0^{(+)}$}

\noindent
We refer to this solution as a ``kink".  Its mirror image
is also a valid solution,
\bq {\widehat\l}_0^{(-)}(t;t_1)=\left\{
   \begin{array}{ll}
    +A & ;\hspace{.2in} t<t_1-\frac{\pi}{2\w} \\
    -A\sin\w(t-t_1) & ;\hspace{.2in}
    t_1-\frac{\pi}{2\w}\le t\le t_1+\frac{\pi}{2\w} \\
    -A & ;\hspace{.2in} t>t_1+\frac{\pi}{2\w} \end{array}\right.,
 \label{lop2}\eq
It
describes an eigenvalue which rolls
from $+A$ to $-A$.
It is referred
to as an ``anti-kink" and
is shown pictorially in figure 4.
\ \vspace{1pc}

\epsfxsize=250pt
\centerline{\epsfbox{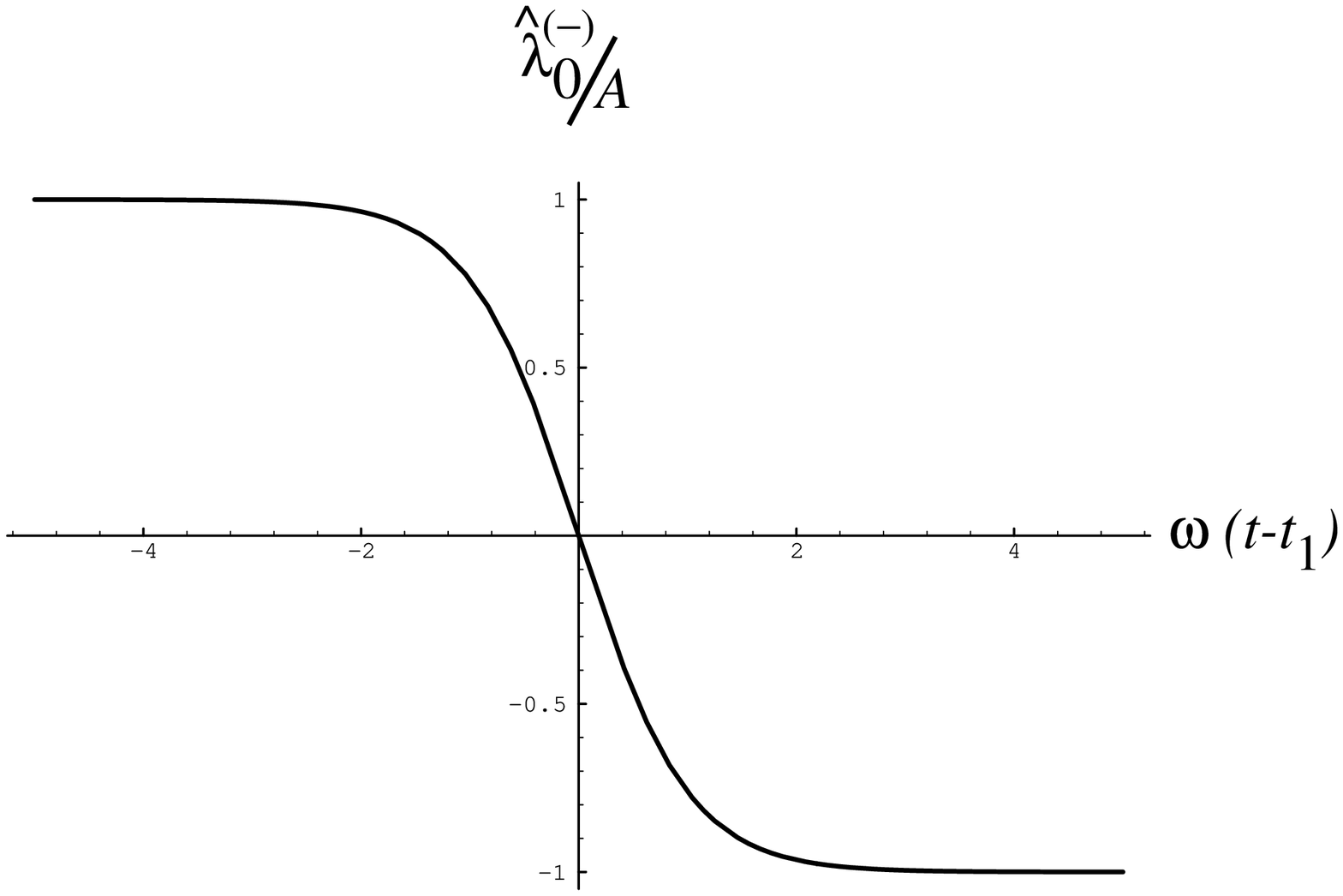}}
\centerline{\tenrm Figure~4. The ``antikink" solution,
${\widehat\l}_0^{(-)}$}

Before discussing more general solutions, it is necessary that we make
a few
clarifying remarks.  As discussed above, when $\l_0=\pm A$ it is
absorbed
into the continuum and does not have a distinct identity.  In the kink
solution $\widehat{\l}_0^{(+)}$ presented above, it is only at
$t=t_1-\frac{\pi}{2\w}$ that the last eigenvalue in the continuum,
located at
$-A$, seperates and leaks into the region between $-A$ and $A$.  At
$t=t_1+\frac{\pi}{2\w}$ the eigenvalue is then reabsorbed into the
continuum at
$+A$.  Similar comments apply to the antikink solution
$\widehat{\l}_0^{(-)}$.
As we will see, it is useful to rephrase these solutions in such a way
that $\l_0$ only exists during the interval
$t_1-\frac{\pi}{2\w}\le t\le t_1+\frac{\pi}{2\w}$.
That is, during the intervals $t<t_1-\frac{\pi}{2\w}$ and
$t>t_1+\frac{\pi}{2\w}$ we refrain from calling any eigenvalue $\l_0$,
since all eigenvalues are then a part of the continuum collective field
$\vphi$.  We therefore rewrite the
kink and antikink solutions as follows,
\bq {\l}_0^{(\pm)}=\pm A\sin\w(t-t_1)
    \hspace{.2in} ; \hspace{.1in}
    t_1-\frac{\pi}{2\w}\le t\le t_1+\frac{\pi}{2\w},
 \label{klink}\eq
which we depict graphically in Figure 5.
\ \vspace{1pc}

\epsfxsize=400pt
\centerline{\epsfbox{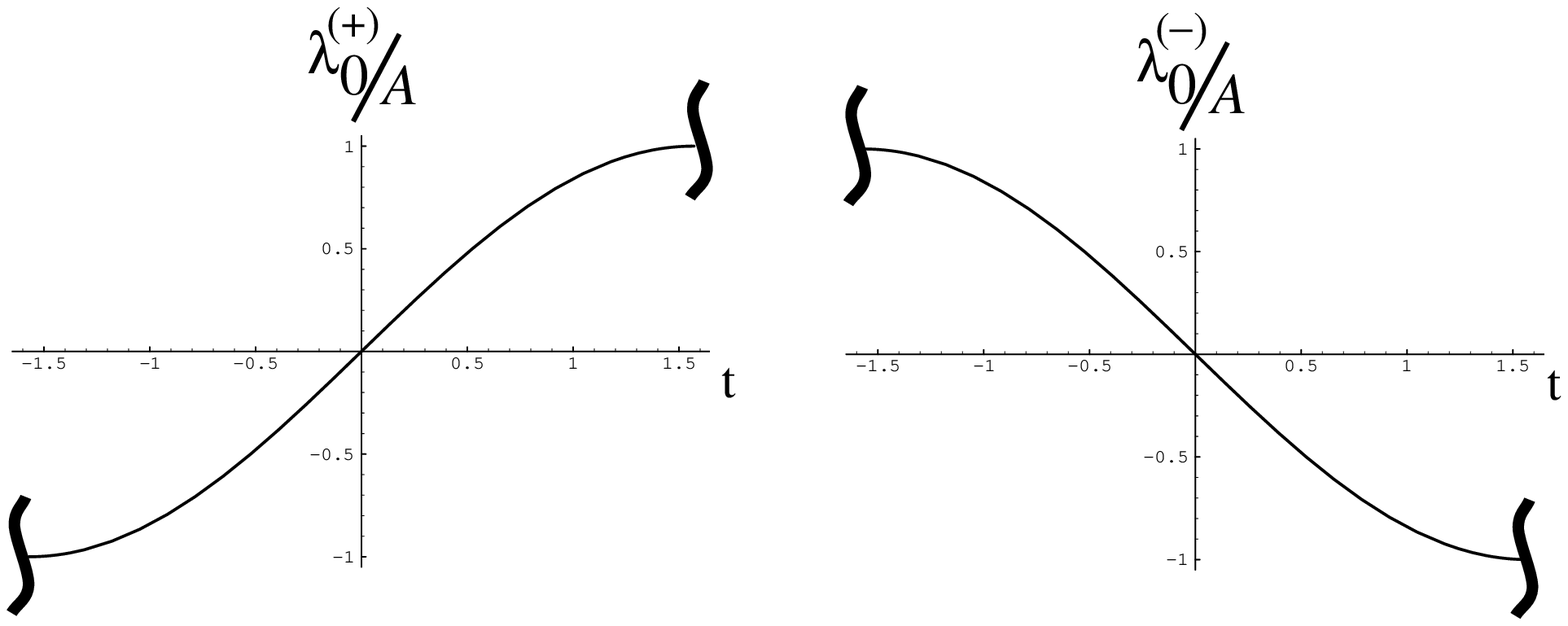}}
\centerline{\tenrm Figure~5. The modified ``kink" and ``antikink"
solutions,
${\l}_0^{(\pm)}$. $t$ is in units in which $\frac{\pi}{2\w}=1$.}

In these graphical representations, the bars at the ends of the kinks
and
antikinks symbolize the emission or reabsorption of
the eigenvalue into the continuum.
The reason why we make this refinement will become clear presently.

There exist more general solutions than those which we have already
discussed, in which the identity of $\l_0$ is a more complex and subtle
issue.  It is possible, for example,
that a kink, which ends with
eigenvalue $\l_0$ attaching to the continuum at $+A$, could be
followed,
at some  later time, by an antikink, in which the eigenvalue
$\l_0$ separates from the continuum at $+A$, rolls to $-A$ and then
reattaches
there.  Such a kink-antikink sequence, which we denote $\l_0^{(+-)}$,
would
satisfy the Euclidean equation of motion,  (\ref{keq}).  It is also
possible,
however, that a kink,
which ends with   the eigenvalue $\l_0$ attaching to the continuum at
$+A$,
could be followed, at some  later time, by another kink in which a
different
eigenvalue detaches from the continuum at $-A$, traverses the region
between
$-A$ and $+A$, and then reattaches to the continuum at $+A$ immediately
next to the eigenvalue involved in the first kink.
This kink-kink sequence, which we denote $\l_0^{(++)}$, also satisfies
(\ref{keq}).  There are thus $2^2=4$ solutions which involve two
distinct
kinks,
\brr {{\l}}_0^{(++)}
   &=&
    \left\{\begin{array}{ll}
    +A\sin\w(t-t_1) &
    ;\hspace{.2in} t_1-\frac{\pi}{2\w}\le t\le t_1+\frac{\pi}{2\w} \\
    +A\sin\w(t-t_2) &
    ;\hspace{.2in} t_2-\frac{\pi}{2\w}\le t\le t_2+\frac{\pi}{2\w}
    \end{array}\right. \nonumber \\
   {\l}_0^{(+-)}
   &=&
    \left\{\begin{array}{ll}
    +A\sin\w(t-t_1) &
    ;\hspace{.2in} t_1-\frac{\pi}{2\w}\le t\le t_1+\frac{\pi}{2\w} \\
    -A\sin\w(t-t_2) &
    ;\hspace{.2in} t_2-\frac{\pi}{2\w}\le t\le t_2+\frac{\pi}{2\w}
    \end{array}\right. \nonumber \\
   {\l}_0^{(-+)}
   &=&
    \left\{\begin{array}{ll}
    -A\sin\w(t-t_1) &
    ;\hspace{.2in} t_1-\frac{\pi}{2\w}\le t\le t_1+\frac{\pi}{2\w} \\
    +A\sin\w(t-t_2) &
    ;\hspace{.2in} t_2-\frac{\pi}{2\w}\le t\le t_2+\frac{\pi}{2\w}
    \end{array}\right. \nonumber \\
   {\l}_0^{(--)}
   &=&
    \left\{\begin{array}{ll}
    -A\sin\w(t-t_1) &
    ;\hspace{.2in} t_1-\frac{\pi}{2\w}\le t\le t_1+\frac{\pi}{2\w} \\
    -A\sin\w(t-t_2) &
    ;\hspace{.2in} t_2-\frac{\pi}{2\w}\le t\le t_2+\frac{\pi}{2\w}
    \end{array}\right.
 \label{lopp}\err
In all four cases
$t_2\ge t_1+\frac{\pi}{\w}$, but both $t_1$ and $t_2$ are otherwise
arbitrary.
We depict the four solutions (\ref{lopp}) graphically in Figure 6.
\ \vspace{1pc}

\epsfxsize=400pt
\centerline{\epsfbox{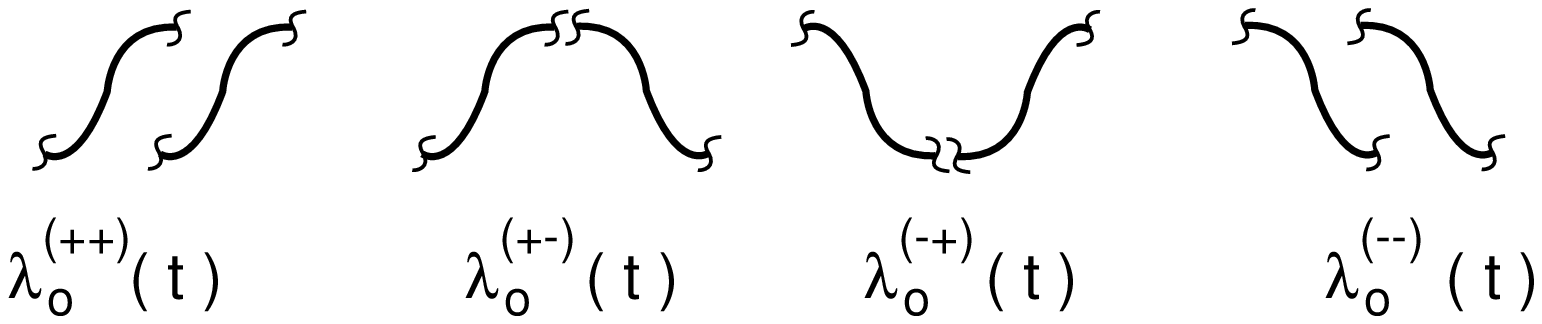}}
\centerline{\tenrm Figure~6. The four solutions (\ref{lopp}).}
\noindent
One might ask whether it is possible for the
second eigenvalue to detach from the continuum while the first
eigenvalue is
still discrete; that is for $t_1\le t_2<t_1+\frac{\pi}{\w}$.
In fact, there do exist
solutions in which two eigenvalues detach
from the
same side of the continuum in quick succession; that is, within a time
interval
less than $\frac{\pi}{\w}$.  The existence of such solutions and their
exact form is actually inconsequential.  This is because
the probability of such a sequence is further suppressed by the
instanton density, which is proportinal to $e^{-\frac{\pi}{2 g}}$. For
sufficiently small $g$, this
probability is negligibly small
and we may therefore consistently ignore these solutions.
We therefore
remove the restriction on $t_1$, $t_2$.
An arbitrary solution consists of $q$ events which are randomly
distributed between kinks and antikinks, where $0\le q<\infty$.
We refer to any such $q$-event solution as a $q$-instanton.  For a
given $q$ there are $2^q$ distinct instanton configurations.
For example, for $q=3$, one solution consists of three consectutive
kinks, which we denote $\l_0^{(+++)}$.  A solution which consists of
a kink followed by two antikinks is denoted $\l_0^{(+--)}$.  Clearly,
for $q=3$, there are $2^3=8$ such solutions.  Generically, we denote
the $2^q$ q-instantons as $\l_0^{(q)}$.  There are $q$ zero modes
associated with each $\l_0^{(q)}$.  These correspond to the arbitrary
times
$t_1,...,t_q$,
where $t_q\ge t_{q-1}+\frac{\pi}{\w}
\cdots\ge t_1+\frac{\pi}{\w}$, when the kinks or antikinks occur.
Once again, we ignore all cases where several eigenvalues are
simultaneously discrete, since the effect of these solutions is
negligible.

This concludes our analysis of the discrete eigenvalue solutions.
In the next section, we take these solutions
as background solutions which we expand around when performing the
path integral associated with the theory.  We integrate the instantons
out of
the path integral and arrive at an
effective
theory  for the collective field $\vphi$ which has the instanton
effects
incorporated explicitly in terms of induced operators.

\renewcommand{\theequation}{3.\arabic{equation}}
\setcounter{equation}{0}
{\fl{\bf 3. Integration Over Instantons}}

The partition function associated with the theory discussed above can
be
written as a sum over different $q$-instanton sectors,
\bq Z=\sum_{q=0}^\infty Z_q
 \label{kashi}\eq
where, schematically,
\bq Z_q=\int[d\vphi]\int[d\l_o]_qe^{-S[\l_0;{\vphi}]}.
 \label{schem}\eq
In this expression
the symbol $[d\l_0]_q$ indicates that $\l_0$ is expanded around
$\l_0^{(q)}$.
For notational convenience we have suppressed a subscript $E$
on the action, but it is assumed throughout this section that we
are in euclidean space.
We proceed to define equation
(\ref{schem}) in more precise terms.  First of all, remember that
$\l_0^{(q)}$
generically represents all the $2^q$ instanton solutions which each
have
$q$ single eigenvalue kinks-antikinks.  Therefore, more specifically,
\bq Z_q=\sum_{\{k_i\}}Z_{k_1\cdots k_q},
 \label{bukhara}\eq
where $k_i=\pm$, the summation is over all $2^q$
possible sets $\{k_1\cdots k_q\}$, and
\bq Z_{k_1\cdots k_q}=\int[d{\vphi}]\int[d\l_0]_{k_1\cdots k_q}
    e^{-S[\l_0;{\vphi}]}.
 \eq
The symbol $[d\l_0]_{k_1\cdots k_q}$ indicates that $\l_0$ is expanded
around $\l_0^{(k_1\cdots k_q)}$ defined in section 2.  Thus,
$Z_2=Z_{++}+Z_{+-}+Z_{-+}+Z_{--}$, and so on.  In order to clarify the
remaining factors in (\ref{schem}) we will focus on an example.

{\fl{\it 3.1 Calculation of $Z_+$}}\\
\indent
Consider $Z_+$, the contribution to the partition function coming from
single
kink configurations.  The correct expression is given by
\bq Z_+=\int[d{\vphi}]\int[d\l_0]_+e^{-S[\l_0;{\vphi}]},
 \label{zeep}\eq
where, as discussed in section 2, we expand $\l_0(t)$ around the
solution
$\widehat{\l}_0^{(+)}(t;t_1)$, which we repeat here for convenience,
\bq \widehat{\l}_0^{(+)}(t;t_1)=\left\{
   \begin{array}{ll}
    -A & ;\hspace{.2in} t<t_1-\frac{\pi}{2\w} \\
    +A\sin\w(t-t_1) & ;\hspace{.2in}
    t_1-\frac{\pi}{2\w}\le t\ge t_1+\frac{\pi}{2\w} \\
    +A & ;\hspace{.2in} t>t_1+\frac{\pi}{2\w} \end{array}\right..
 \eq
The parameter $t_1$ is
a zero mode of this solution, and an integration over $t_1$ is implied
within $\int[d\l_0]_+$.  To make this explicit, we write
\bq \l_0(t)=\widehat{\l}_0^{(+)}(t;t_1)+\tilde{\l}(t).
 \eq
Thus, extracting a $t_1$ integration,
we obtain
\bq \int[d\l_0]_+=\sqrt{\frac{\pi}{2 g}}\cdot\int_{-\infty}^\infty
    dt_1\int[d\tilde{\l}']
 \label{cluge}\eq
where $\int[d\tilde{\l}']$ indicates integration over all functions
orthogonal to
$\dot{\widehat{\l}}_0^{(+)}$;
that is, all functions $\tilde{\l}'(t)$ such that
\bq \int_{-\infty}^\infty
dt\tilde{\l}'(t)\dot{\widehat{\l}}_0^{(+)}(t;t_1)=0.
 \label{oort}\eq
The factor $\sqrt{\frac{\pi}{2g}}$ in (\ref{cluge}) is a jacobian.
We may now rewrite equation (\ref{zeep}) as follows,
\brr Z_+ &=& \sqrt{\frac{\pi}{2g}}\int dt_1
     \int[d{\vphi}]\int[d\tilde{\l}']
     e^{-S[\l_0;{\vphi}]}.
 \label{klib}\err
Recall that when $t<t_1-\frac{\pi}{2\w}$ and when
$t>t_1+\frac{\pi}{2\w}$ the eigenvalue $\l_0$ is not discrete,
but is actually part of the continuum.
We therefore define $\widehat{\vphi}$ as the
continuous collective field with $N+1$ eigenvalues $\l_0,\l_1,...\l_N$,
in order to distinguish it from $\vphi$, which
has only $N$ eigenvalues, $\l_1,...\l_N$. The limit
$N\rightarrow\infty$
is assumed in both cases.
Hence, when $t<t_1-\frac{\pi}{2\w}$ and when
$t>t_1+\frac{\pi}{2\w}$ we can
write
$\int[d{\vphi}]\int[d\tilde{\l}']=\int[d\widehat{\vphi}]$ and
$S[\l_0;{\vphi}]=S_\vphi[\widehat{\vphi}]$,
where $S_\vphi[\widehat{\vphi}]$ is the euclidean collective field
action
given in equation (2.8), here expressed as a function of
$\widehat{\vphi}$
rather than $\vphi$.
Thus, $Z_+$ can be factored as
\brr Z_+ &=& \sqrt{\frac{\pi}{2g}}\int dt_1\bl
     \int[d{\widehat{\vphi}}]
     \exp\bbl{-S_\vphi[\widehat{\vphi}]
     \bb_{-\infty}^{t_1-\frac{\pi}{2\w}}}\bbr \nonumber\\
     & & \hspace{.8in} \times
     \int[d{\vphi}]\int[d\tilde{\l}']
     \exp\bbl{-S[\l_0;{\vphi}]\bb_{t_1-\frac{\pi}{2\w}}
     ^{t_1+\frac{\pi}{2\w}}}\bbr \nonumber\\
     & & \hspace{.8in} \times
     \int[d{\widehat{\vphi}}]
     \exp\bbl{-S_\vphi[\widehat{\vphi}]\bb_{t_1+\frac{\pi}{2\w}}
     ^\infty}\bbr\br.
 \label{kliby}\err
where $S\bb_a^b=\int_a^b dt L$ and the functional integrals cover
functions
defined only during the time intervals specified in the associated
integrands.
It is useful to convert the remaining
$\int [d{\vphi}]$ integration in (\ref{kliby}) into a
$\int [d\widehat{\vphi}]$ integration.  We proceed to do this.
Let us denote the middle factor in (\ref{kliby}) by
\bq z_+(t_1)=\int[d{\vphi}]\int[d\tilde{\l}']\exp\bl-S[\l_0;{\vphi}]
    \bb_{t_1-\frac{\pi}{2\w}}^{t_1+\frac{\pi}{2\w}}\br,
 \label{zep}\eq
where $\l_0(t)=\l_0^{(+)}(t;t_1)+\tilde{\l}'(t)$,
$t_1$ is taken as a constant parameter, and from (2.21) we have
\bq S[\l_0;{\vphi}]
    =S_\vphi[{\vphi}]+\int dt\bl
    \hf\dot{\l}_0^2+\hf\w^2(A^2-\l_0^2)
    +\int dx\frac{{\vphi}'}{(x-\l_0)^2}\br.
 \label{shed}\eq
We now consider the following representation of the number $1$,
\bq 1=\frac{\int[d\l_{\so}]_+\exp\bl -S_{\so}
    [\l_{\so};{\vphi}]
    \bb_{t_1-\frac{\pi}{2\w}}
     ^{t_1+\frac{\pi}{2\w}}\br}
    {\int[d\l_{\so}]_+
    \exp\bl -S_{\so}[\l_{\so};{\vphi}]
    \bb_{t_1-\frac{\pi}{2\w}}^{t_1+\frac{\pi}{2\w}}\br},
 \label{one}\eq
where
\bq S_{\so}[\l_{\so};{\vphi}]
     = \int dt
     \bl\hf\dot{\l}_{\so}^2+\hf\w^2(A^2-\l_{\so}^2)
     +\int dx\frac{{\vphi}'}{(x-\l_{\so})^2}\br
 \label{sop}\eq
and $\l_{\so}$ is a dummy eigenvalue
expanded around a particular background
$\l_{\so}^{(+)}$ given by
\bq \l_{\so}^{(+)}(t;t_1)=\left\{\begin{array}{ll}
      -A & ;\hspace{.2in} t_1-\frac{\pi}{2\w}\le t<t_1 \\
      +A &  ;\hspace{.2in} t_1<t\le t_1+\frac{\pi}{2\w}
\end{array}\right..
 \label{step}\eq
Thus, we define
\bq \l_{\so}(t)=\l_{\so}^{(+)}(t;t_1)+\widehat{\l}(t).
 \eq
Taylor expanding (\ref{sop}) in $\widehat{\l}$, we find
\bq S_{\so}[\l_{\so};{\vphi}]
      =\int dt \bl
      \int dx\frac{{\vphi}'}{(x-\l_{\so}^{(+)})^2}
      +\hf\widehat{\l}\O_0\widehat{\l}+{\cal O}(\widehat{\l}^3)\br,
 \label{tajik}\eq
where
\bq \O_0=-\der_t^2-\w^2+\int dx
    \frac{6\vphi'}{(x-\l_{\so}^{(+)})^4}.
 \label{oodef}\eq
Note that, to lowest order in $\e_x$, the contribution linear in
$\widehat{\l}$
vanishes.  This is because $\l_{\so}=\pm A$ satisfies the
static equation of motion derived from (\ref{sop}) to order $\e_x^0$,
as
discussed in the paragraph which follows equation (2.29).
In this paper we will restrict ourselves to the semi-classical
approximation.  In this approximation we may replace ${\vphi}'$ in
operator
$\O_0$
with $\tilde{\vphi}_0^{'}$ given in equation (2.11).  Furthermore,
since
$t_1$ is a constant, it follows that $[d\l_{\so}]_+=[d\widehat{\l}]$.
It is then straightforward to show that
\bq \int[d\l_{\so}]_+
    \exp\bl -S_{\so}[\l_{\so};{\vphi}]
    \bb_{t_1-\frac{\pi}{2\w}}^{t_1+\frac{\pi}{2\w}}\br
    =\frac{1}{\sqrt{\det{\O_0}}}
    \exp\bl
    -\int_{t_1-\frac{\pi}{2\w}}^{t_1+\frac{\pi}{2\w}}dt
    \int dx
    \frac{{\vphi}'}{(x-\l_{\so}^{(+)})^2}\br
 \label{biz}\eq
where
\bq
\det\O_0=\bbl\int[d\widehat{\l}]\exp\bl-\hf\int_{t_1-\frac{\pi}{2\w}}
    ^{t_1+\frac{\pi}{2\w}} dt
    \widehat{\l}\O_0\widehat{\l}\br\bbr^{-2}.
 \label{ddet}\eq
Inserting (\ref{biz}) in the denominator of (\ref{one}) we find that
\bq 1 = \sqrt{\det{\O_0}}
    \int[d\l_{\so}]_+\exp\bl -S_{\so}
    [\l_{\so};{\vphi}]
    \bb_{t_1-\frac{\pi}{2\w}}
     ^{t_1+\frac{\pi}{2\w}}+
    \int_{t_1-\frac{\pi}{2\w}}^{t_1+\frac{\pi}{2\w}}dt
    \int dx
    \frac{{\vphi}'}{(x-\l_{\so}^{(+)})^2}\br.
 \label{onep}\eq
Inserting (\ref{onep}) into (\ref{zep}), we then obtain
\brr z_+(t_1) &=&
     \sqrt{\det\O_0}
     \int[d{\vphi}]\int[d\l_{\so}]_+\int[d\tilde{\l}']
     \nonumber \\
      & & \hspace{.1in} \times
     \exp\bl-({S}[\l_0;{\vphi}]+S_{\so}[\l_{\so};{\vphi}])
     \bb_{t_1-\frac{\pi}{2\w}}^{t_1+\frac{\pi}{2\w}}
     +\int_{t_1-\frac{\pi}{2\w}}^{t_1+\frac{\pi}{2\w}}dt
     \int dx
     \frac{{\vphi}'}{(x-\l_{\so}^{(+)})^2}\br.
 \label{ninc}\err
Now, using (\ref{shed}) and (\ref{sop}) we find
\brr {S}[\l_0;{\vphi}]+S_{\so}[\l_{\so};{\vphi}]
     &=&
     S_\vphi[{\vphi}]+
     \int dt
     \bl\hf\dot{\l}_{\so}^2+\hf\w^2(A^2-\l_{\so}^2)
     +\int dx\frac{{\vphi}'}{(x-\l_{\so})^2}\br \nonumber \\
     & & \hspace{.3in}
     +\int dt\bl\hf\dot{\l}_0^2+\hf\w^2(A^2-\l_0^2)
     +\int dx\frac{{\vphi}'}{(x-\l_0)^2}\br.
 \label{stst}\err
However, rewriting the collective field ${\vphi}$  in terms of
eigenvalues
the first two terms of the last expression may be rewritten as follows,
\brr & & S_\vphi[{\vphi}]+
     \int dt
     \bl\hf\dot{\l}_{\so}^2+\hf\w^2(A^2-\l_{\so}^2)
     +\int dx\frac{{\vphi}'}{(x-\l_{\so})^2}\br \nonumber \\
     & & \hspace{.4in}
     =\sum_{i=1}^N\bl\hf\dot{\l}_i^2+\hf\w^2(A^2-\l_i^2)
     +\hf\sum_{j\ne i,\so}\frac{1}{(\l_j-\l_i)^2}\br\nonumber \\
     & & \hspace{.76in}
     +\hf\dot{\l}_{\so}^2+\hf\w^2(A^2-\l_{\so}^2)
     +\sum_{j\ne \so}\frac{1}{(\l_j-\l_{\so})^2} \nonumber \\
     & & \hspace{.4in}
     =\int dt\sum_{i=\so}^N\bl
     \hf\dot{\l}_i^2+\hf\w^2(A^2-\l_i^2)
     +\hf\sum_{j\ne i}\frac{1}{(\l_j-\l_i)^2}\br \nonumber \\
     & & \hspace{.4in}
     =S_\vphi[\widehat{\vphi}].
 \label{st}\err
Thus, $\l_{\so}$ replaces the missing eigenvalue in ${\vphi}$.
Note that $\l_{\so}$ is expanded around (\ref{step})
which is exactly the classical configuration of the missing eigenvalue.
It follows that, over the range $t_1-\frac{\pi}{2\w}\le t\le
t_1+\frac{\pi}{2\w}$,
\bq \int[d{\vphi}]\int[d\l_{\so}]_+=\int[d\widehat{\vphi}].
 \label{crud}\eq
Using  (\ref{stst}), (\ref{st}) and (\ref{crud})
in equation (\ref{ninc}) we obtain
\brr z_+(t_1) &=& \sqrt{\det\O_0}\int[d\widehat{\vphi}]
     \exp\bl-S_\vphi[\widehat{\vphi}]
     \bb_{t_1-\frac{\pi}{2\w}}^{t_1+\frac{\pi}{2\w}}\br
      \nonumber \\
     & & \hspace{.4in}\times
    \int[d\tilde{\l}']\exp\bl
    -(S_0[\l_0]+S_I[\l_0;{\vphi}])
    \bb_{t_1-\frac{\pi}{2\w}}^{t_1+\frac{\pi}{2\w}}\br,
 \label{almo}\err
where
\bq S_0[\l_0]
    =\int dt
    \bl\hf\dot{\l}_0^2+\hf\w^2(A^2-\l_0^2)\br
 \label{sin}\eq
and
\bq S_I[\l_0;{\vphi}]=\int dt\int dx\bl
     \frac{{\vphi}'}{(x-\l_0)^2}
    -\frac{{\vphi}'}{(x-\l_{\so}^{(+)})^2}\br.
 \label{bad}\eq
We would now like to perform the $\int[d\tilde{\l}']$ integration in
(\ref{almo}).  First, recall that
$\l_0(t)=\l_0^{(+)}(t;t_1)+\tilde{\l}(t)$.
Inserting this into (\ref{sin}) and (\ref{bad}), we obtain
\bq S_0[\l_0]+S_I[\l_0;{\vphi}] =
    S_0[\l_0^{(+)}]+S_I[\l_0^{(+)};{\vphi}]
    +\hf\int dt\tilde{\l}\O_1\tilde{\l}
    +\O(\tilde{\l}^3),
 \eq
where
\bq \O_1=-\der_t^2-\w^2+\int dx\frac{6\vphi'}{(x-\l_0^{(+)})^4}.
 \label{oidef}\eq
As before, the term linear in $\tilde{\l}$ vanishes since $\l_0^{(+)}$
is a solution of the equation of motion to order $\e_x^0$.
Furthermore, since we want to work to lowest order in $\hbar$ only, we
may replace $\vphi'$ in operator $\O_1$ by $\tilde{\vphi}_0^{'}$ given
in
(\ref{vback}).
Therefore
\brr & &\int[d\tilde{\l}']\exp
    \bl-(S_0[\l_0]+S_I[\l_0;\vphi])
    \bb_{t_1-\frac{\pi}{2\w}}^{t_1+\frac{\pi}{2\w}}\br \nonumber\\
    &=& \hspace{.2in} \frac{1}{\sqrt{\det'\O_1}} \exp
    \bl-(S_0[\l_0^{(+)}]+S_I[\l_0^{(+)};\vphi])
    \bb_{t_1-\frac{\pi}{2\w}}^{t_1+\frac{\pi}{2\w}}\br
 \err
where
\bq {\det}'\O_1=\bbl\int[d\tilde{\l}']\exp\bl-\hf
    \int_{t_1-\frac{\pi}{2\w}}^{t_1+\frac{\pi}{2\w}}dt_1
    \tilde{\l}\O_1\tilde{\l}\br\bbr^{-2}.
 \eq
The action of the single-kink instanton is easily evaluated.
It is given by
\bq S_0[\l_0^{(+)}]\bb_{t_1-\frac{\pi}{2\w}}^{t_1+\frac{\pi}{2\w}}
    =\frac{\pi}{2g}.
 \eq
Using the above results in (\ref{almo})
we find
\bq z_+(t_1)=
   \sqrt{\frac{\det\O_0}{\det'\O_1}}  \hbox{\large $e$}^
{\hbox{$-\frac{\pi}{2g}$} }
    \int_{-\infty}^{\infty}dt_1
    \int[d\widehat{\vphi}]\exp\bl
    -(S_\vphi[\widehat{\vphi}]+S_I[\l_0^{(+)};{\vphi}])
    \bb_{t_1-\frac{\pi}{2\w}}^{t_1+\frac{\pi}{2\w}}\br.
 \label{yak}\eq
Using equations (\ref{yak}) and (\ref{zep}) in equation (\ref{kliby})
then yields
\brr Z_+ &=& \M\int_{-\infty}^\infty dt_1\int[d\widehat{\vphi}]\exp
\bl-S_\vphi[\widehat{\vphi}]\bb_{-\infty}^\infty\br \nonumber \\
& &\hspace{1in}\times\exp\bl-S_I[\l_0^{(+)};\vphi]
\bb_{t_1-\frac{\pi}{2\w}}^{t_1+\frac{\pi}{2\w}}\br,
\label{kurd}\err
where
\bq \M=\sqrt{\frac{\pi}{2g}}
\sqrt{\frac{\det\O_0}{\det'\O_1}} e^{-\frac{\pi}{2g}}.
\label{mdef}\eq
In Appendix A we explicitly compute this quantity.  We find,
for ``reasonable" values of $g$, that
$\M\approx {\w}\sqrt{\frac{\pi}{2 g}}
\hbox{\large $e$}^ {\hbox{$-\frac{\pi}{2g}$} }$
where the constant of proportionality is  $\O(1)$.  In the remainder of
this paper, for simplicity, we will set this constant to unity.
In this case
\bq \M={\w}\sqrt{\frac{\pi}{2 g}}
\hbox{\large $e$}^ {\hbox{$-\frac{\pi}{2g}$} }
{}.
 \label{madef}\eq
The scale $\M$ is an important quantity since it sets the scale of all
nonperturbative effects in the theory.
At this point, the distinction between $\widehat{\vphi}$ and $\vphi$
becomes immaterial, so we will henceforth omit the hat on
$\widehat{\vphi}$.
We will also suppress the limits $-\infty$ and $\infty$ on
$S_\vphi[\vphi]$
and on the $\int dt_1$ integration, and we will abbreviate
$S_I$ as follows
\bq  S_I[\l_0^{(+)};\vphi]
    \bb_{t_1-\frac{\pi}{2\w}}^{t_1+\frac{\pi}{2\w}}
    =S_I^{(+)}[\vphi;t_1].
 \eq
Equation (\ref{kurd}) can then be rewritten more concisely as
\bq Z_+=\M\int
dt_1\int[d\vphi]e^{-S_\vphi[\vphi]}e^{-S_I^{(+)}[\vphi;t_1]}.
 \eq
This concludes the example calculation of $Z_+$.

{\fl{\it 3.2 Calculation of $Z$}}\\
\indent
When we perform the same analysis on an arbitrary $Z_{k_1\cdots k_q}$,
as
we did above on the case $Z_+$, we arrive at the following general
result,
\bq Z_{k_1\cdots k_q}=\M^q\prod_{i=1}^q
    \int_>dt_i\int[d\vphi]e^{-S_\vphi[\vphi]}
    \prod_{j=1}^qe^{-S_I^{(k_j)}[\vphi;t_j]},
 \label{peshwar}\eq
where $k_i=\pm$, $\prod_i\int_>dt_i$ is an ordered set of nested
integrals,
\bq \prod_{i=1}^q\int_>dt_i=\int_{-\frac{T}{2}}^\frac{T}{2}dt_1
\int_{t_1+\frac{\pi}{\w}}^{\frac{T}{2}}dt_2
\cdots\int_{t_{q-1}+\frac{\pi}{\w}}^{\frac{T}{2}}dt_q,
\label{nest}\eq
where $T\rightarrow\infty$,
and $S_I^{(\pm)}$ is a generalization of (\ref{bad}),
\bq
S_I^{(\pm)}[\vphi;t_j]=
\int_{t_j-\frac{\pi}{2\w}}^{t_j+\frac{\pi}{2\w}}dt
\int dx\bl\frac{\vphi'(x,t)}{(x-\l_0^{(\pm)}(t-t_j))^2}
-\frac{\vphi'(x,t)}{(x-\l_{\so}^{(\pm)}(t-t_j))^2}\br.
 \label{dali}\eq
Using
(\ref{kashi}), (\ref{bukhara})
and (\ref{peshwar}) we now find that
\bq Z=\int[d\vphi]e^{-S_\vphi[\vphi]}\sum_{q=0}^\infty
    \M^q\prod_{i=1}^q\int_>dt_i\sum_{\{k_i\}}\prod_{j=1}^q
    e^{-S_I^{(k_j)}[\vphi;t_j]}.
 \label{lhasa}\eq

Notice that we
do not let any pair of $t_i$'s come within $\D t=\frac{\pi}{\w}$ of
each other.
As discussed in section 2,
the reason
for this is that the probability
for configurations in which any pair of $t_i$'s are
within this range is negligibly
small. Such configurations, in which two kinks or
anti-kinks
overlap include complicated instanton-instanton interactions. As
in more traditional
instanton calculations these interactions are of strength
proportional to the square of the instanton fugacity,
$(e^{-\frac{\pi}{2g}})^2$ and therefore add negligible correction. We
will
therefore ignore them. This is
a dilute gas approximation. The practical consequence of this
is to remove the restriction on the range of the $t_i$'s.
The integrand is then
completely symmetric under
$t_i\leftrightarrow t_j$ for any $i$ and $j$. We may therefore replace
the
ordered (and unrestricted) $\int_>dt_i$ integrals
with unordered integrals provided we insert a factor
of $1/q!$ to compensate for overcounting.
  Thus, we may rewrite
(\ref{lhasa}) as follows,
\brr Z &=& \int[d\vphi]e^{-S_\vphi[\vphi]}\sum_{q=0}^\infty
     \frac{1}{q!}\M^q\prod_{i=1}^q
     \int dt_i\sum_{\{k_i\}}\prod_{j=1}^q
     e^{-S_I^{(k_j)}[\vphi;t_j]} \nonumber \\
     &=& \int[d\vphi]e^{-S_\vphi[\vphi]}\sum_{q=0}^\infty
     \frac{1}{q!}
     \bl\M\int dt_1\bpl e^{-S_I^{(+)}[\vphi;t_1]}
     +e^{-S_I^{(-)}[\vphi;t_1]}\bpr\br^q.
 \err
The sum over $q$ is now an exponential, so that
\bq Z=\int[d\vphi]e^{-S_{eff}[\vphi]},
 \eq
where
\bq S_{eff}[\vphi]=S_\vphi[\vphi]+\Delta S[\vphi]
 \eq
is the effective action with the $q$-instanton effects systematically
incorporated, and
\bq \Delta S[\vphi]=\M\int dt_1
    \bl e^{-S_I^{(+)}[\vphi;t_1]}
     +e^{-S_I^{(-)}[\vphi;t_1]}\br
 \label{delta}\eq
is the associated change in the action, where $\M$ and $S_I^{(\pm)}$
are given in (\ref{mdef}) and (\ref{dali}) respectively.
Equation (\ref{delta}) is the change in the collective field action
due to the presence of $q$-instantons, in the limit of small $g$.
Note that this expression is not a two dimensional integral over a
local
density.

We should express the collective field theory, and any
instanton-induced operators, in terms of canonically propagating
fields.  We begin the following subsection with the identification of
the
canonical theory, and proceed to re-analyze the above results in this
more
appropriate framework.

{\fl{\it 3.3 Canonical Theory }}\\
\indent
So far in this paper we have studied the collective field theory
expressed in terms of the field $\vphi$.  By examining equation
(\ref{se}),
however, we discover that $\vphi$ does not have a canonically
normalized kinetic energy.  We also find that the collective field
Lagrangian is neither Lorentz invariant nor translationally invariant.
The first of these problems is solved, in part, by expanding $\vphi$
around
the solution to the euclidean field equation $\tilde{\vphi}_0$ given in
(\ref{vback}).  Thus, we define
\bq \vphi(x,t)=\tilde{\vphi}_0(x)+\frac{1}{\sqrt{\pi}}\z(x,t).
 \eq
As discussed at length elsewhere, a canonical kinetic energy
is obtained by expressing the Lagrangian in terms of
a new spatial coordinate $\tau$ defined by the following
relation,
\bq \tau'(x)=\frac{1}{\pi}(\tilde{\vphi}_0^{'}(x))^{-1}.
 \label{tox}\eq
Note that $\tau$ has mass dimension $-1$, which is the appropriate mass
dimension for a spatial coordinate, whereas $x$ has mass dimension
$-\hf$.  Expressing the euclidean collective field action (\ref{se}) in
terms of $\z(\tau,t)$, we find, in the absence of instanton effects,
that
\bq S_\z[\z] = \int dt\int d\tau\bl
       \hf(\dot{\z}^2+\z^{'2})
       -\hf\frac{\gbig(\tau)\dot{\z}^2\z'}{1+\gbig(\tau)\z'}
       +\frac{1}{6}\gbig(\tau)\z^{'3}
       -\frac{1}{3}\frac{1}{\gbig(\tau)^2}\br,
 \label{zest} \eq
where $\gbig(\tau)$ is a space dependent coupling parameter, which we
define below, and the $\tau$ integration is over the limits
$-\infty<\tau\le \tau_0+\frac{\s}{2}$
and $\tau_0+\frac{\s}{2}\le\tau<\infty$,
where $\tau_0$ and $\s$ are independent integration constants which
arise when solving (\ref{tox}).
The reason why there are two integration
constants rather than one, given that (\ref{tox}) is a first-order
differential equation, is that we must solve (\ref{tox}) independently
over
the two seperate regions $-\infty<x\le A$ and $A\le x<\infty$.
The region $-A<x<A$, where there is no continuous
collective field theory, is the low density region. In $\tau$ space,
this region is given by $\tau_0-\frac{\s}{2}<\tau<\tau_0+\frac{\s}{2}$,
so that
$\tau_0$ is the center of the low density region and $\s$ is the
width.
The coupling parameter, defined over
$-\infty<\tau\le \tau_0-\frac{\s}{2}$ and
$\tau_0+\frac{\s}{2}\le\tau<\infty$,
is given by $\gbig(\tau)=(\pi^{3/2}\tilde{\vphi}_0(x))^{-1}$, and
is found to be
\bq \gbig(\tau)=
     4\sqrt{\pi}\frac{g}{\w}\frac{\frac{1}{\k} e^{-2\w|\tau-\tau_0|}}
     {(1-\frac{1}{\k}e^{-2\w|\tau-\tau_0|})^2},
 \label{ggdef}\eq
where $\k$ is a dimensionless number,
\bq \k=\exp(-\w\s),
 \label{kdef}\eq
which relates the width, $\s$, of the low density region in $\tau$
space
to the natural length scale in the matrix model, $1/\w$.
Notice that the coupling
parameter blows up as $\tau\rightarrow\tau_0\pm\frac{\s}{2}$; that is,
at
the boundaries of the low density region.

We would now like to express the change in the effective action due to
the
$q$-instanton effects, equation (\ref{delta}), in terms of the
canonical variable $\z(\tau,t)$.  Since $S_I^{(\pm)}$ is linear in
$\vphi$, it follows that
\bq S_I^{(\pm)}[\vphi;t_1]=S_I^{(\pm)}[\tilde{\vphi}_0]
    +\frac{1}{\sqrt{\pi}}S_I^{(\pm)}[\z;\tau_0,t_1].
 \label{erg}\eq
The $\tau_0$ dependence in the last term of this equation will be
made clear presently.
{}From (\ref{dali}), we find
\bq
S_I^{(\pm)}[\z;\tau_0,t_1]=\int_{t_1-\frac{\pi}{2\w}}^{t_1+
\frac{\pi}{2\w}}dt\int
d\tau\bl\frac{\z'(\tau,t)}{(x(\tau)-\l_0^{(\pm)}(t-t_1))^2}
    -\frac{\z'(\tau,t)}{(x(\tau)-\l_{\so}^{(\pm)}(t-t_1))^2}\br,
 \label{sox}\eq
where the prime now means differentiation with respect to $\tau$,
and where
\bq x(\tau)=\left\{\begin{array}{ll}
    -A\cosh\{\w(\tau-\tau_0+\s/2)\} & \hspace{.2in} ;
\tau\le\tau_0-\s/2 \\
     +A\cosh\{\w(\tau-\tau_0-\s/2)\} & \hspace{.2in} ;
\tau\ge\tau_0-\s/2
     \end{array}\right. .
 \label{xot}\eq
This last expression is found by integrating (\ref{tox}) to obtain
$\tau(x)$ and then inverting the result to obtain $x(\tau)$.
This function depends explicitly on $\tau_0$.  This explains why there
is an explicit $\tau_0$ in equations (\ref{erg}) and (\ref{sox}).
It is straightforward to compute $S_I^{(\pm)}[\tilde{\vphi}_0]$,
given (\ref{vback}), (\ref{dali}), and the definitions
(\ref{klink}) and (\ref{step}).
We emphasize, however, that one must
include the cutoff $\e_x$ in the lower limit of the
$x$ integration in this expression.
We find
\bq S_I^{(\pm)}[\tilde{\vphi}_0]
     = -2^{3/2}\sqrt{\frac{A}{\e_x}}
     +\ln\sqrt{\frac{A}{\e_x}}+\O(\frac{\e_x}{A}).
 \label{ploonp}\eq
As discussed above, $\e_x$ is the size
of the inter-eigenvalue seperation near the edge of the continuum and
so
provides the natural regulator for expressions such as (\ref{ploonp}).
{}From (\ref{edef}) it follows that, to lowest order in $g$,
$e^{-S_I^{(\pm)}[\tilde{\vphi}_0]}\approx g^{1/3}e^{\O(g^{1/3})} $
The constant of proportionality in this expression is $\O(1)$.
In the remainder of this paper, for simplicity, we will set this
constant to unity. In this case
\bq e^{-S_I^{(\pm)}[\tilde{\vphi}_0]}= g^{1/3}e^{\O(g^{1/3})}.
 \label{ploon}\eq
Since all $x$-space integrations are cut-off at a distance $\e_x$ from
the
edge of the low density region; that is, at $|x|=A+\e_x$, it follows
that
all $\tau$ space integrals must be cut-off as well at a value
$\e_\tau$.
Specifically, in (\ref{sox}) and in all other expressions in this paper
which include a $\int d\tau$ integration, the following is implied,
\bq \int d\tau=\int_{-\infty}^{\tau_0-\frac{\s}{2}-\e_\tau}d\tau
    +\int_{\tau_0+\frac{\s}{2}+\e_\tau}^\infty d\tau.
 \eq
The value of $\e_\tau$ is simple to obtain.  We require that
\brr x(\tau-\frac{\s}{2}-\e_\tau) &=& -A-\e_x \nonumber \\
     x(\tau+\frac{\s}{2}+\e_\tau) &=& A+\e_x.
 \err
Using (\ref{xot}) and (\ref{edef}) it follows, to leading order in $g$,
that
\bq \e_\tau=\frac{1}{\w\sqrt{2}}(3\pi g)^{1/3}.
 \eq
Now, using (\ref{ploon}), substituting (\ref{erg}) into (\ref{delta}),
and using (\ref{madef}),
we find that
\bq \Delta S[\z]=\w g^{-1/6}e^{-\frac{\pi}{2g}}\int dt_1
    \bl e^{-S_I^{(+)}[\z;\tau_0,t_1]}
     +e^{-S_I^{(-)}[\z;\tau_0,t_1]}\br.
 \label{zelta}\eq
To recap our results so far, the partition function for the collective
field theory, including the instanton effects, expressed in terms
of the canonical field $\z$ is
\bq Z=\int[d\z]e^{-S_{eff}[\z]},
 \eq
where
\bq S_{eff}[\z]=S_\z[\z]+\D S[\z],
 \eq
$S_\z[\z]$ is given in (\ref{zest}), $\D S[\z]$ is given in
(\ref{zelta}),
and the functions $S_I^{(\pm)}[\z;\tau_0,t_1]$ which appear in
(\ref{zelta}) are given in (\ref{sox}).  Equation (\ref{zelta}) is
a significant result.
Concisely, it is the induced change in the canonical collective field
theory which results from the systematic inclusion of instanton
effects.    As we will demonstrate in the next subsection, equation
(\ref{zelta}) includes operators higher order in $g$.
We will also demonstrate that this result includes nonlocal
interactions. We
will address each of these two issues and conclude the following
subsection by exposing a more useful form for the induced action, as a
two-dimensional integral over a density function expressed
consistently to lowest order in $g$.

{\fl{\it 3.4 Lowest Order Induced Action as an Integral Over a
Local Density  }}\\
We begin by focussing on $S_I^{(\pm)}$.  It is useful to
seperate these into two pieces,
\bq S_I^{(\pm)}=S_{<I}^{(\pm)}+S_{>I}^{(\pm)},
 \eq
where $S_{<I}^{(\pm)}$ includes the contribution coming from
the region $\tau<\tau_0-\frac{\s}{2}$, and $S_{>I}^{(\pm)}$
includes the contribution coming from the region
$\tau>\tau_0+\frac{\s}{2}$.
Using (\ref{dali}), we may write these as follows
\bq S_{<I}^{(\pm)}[\z;\tau_0,t_1]=\int_{t_1-\frac{\pi}{2\w}}
          ^{t_1+\frac{\pi}{2\w}}dt
    \int_{-\infty}^{\tau_0-\frac{\s}{2}-\e_\tau} d\tau
\J_<^{(\pm)}(\tau-\tau_0+\frac{\s}{2},t-t_1)\z'(\tau,t)
\label{sionel} \label{sleft}\eq
and
\bq S_{>I}^{(\pm)}[\z;\tau_0,t_1]=\int_{t_1-\frac{\pi}{2\w}}
          ^{t_1+\frac{\pi}{2\w}}dt
     \int_{\tau_0+\frac{\s}{2}+\e_\tau}^{\infty} d\tau
\J_>^{(\pm)}(\tau-\tau_0-\frac{\s}{2},t-t_1)\z'(\tau,t)
\label{sioner},\label{sright}\eq
where
\bq \J_<^{(\pm)}(\tau-\tau_0+\frac{\s}{2},t-t_1)=
    \frac{1}{(x(\tau-\tau_0+\frac{\s}{2})-\l_0^{(\pm)}(t-t_1))^2}-
\frac{1}{(x(\tau-\tau_0+\frac{\s}{2})-\l_{\emptyset}^{(\pm)}(t-t_1))^2}
 \eq
and
\bq \J_>^{(\pm)}(\tau-\tau_0-\frac{\s}{2},t-t_1)=
    \frac{1}{(x(\tau-\tau_0-\frac{\s}{2})-\l_0^{(\pm)}(t-t_1))^2}-
\frac{1}{(x(\tau-\tau_0-\frac{\s}{2})-\l_{\emptyset}^{(\pm)}(t-t_1))^2}
\eq
By shifting the arguments of the integration in (\ref{sleft}) and
(\ref{sright})
as $\tau\rightarrow\tau+\tau_0\mp\frac{\s}{2}$ respectively
and $t\rightarrow t+t_1$, we then obtain
\bq S_{<I}^{(\pm)}[\z;\tau_0,t_1]
    =\int_{-\frac{\pi}{2\w}}^{\frac{\pi}{2\w}}dt
    \int_{-\infty}^{-\e_\tau} d\tau
    \J_<^{(\pm)}(\tau,t)\z'(\tau+\tau_0-\frac{\s}{2},t+t_1).
 \label{si2l}\eq
and
\bq S_{>I}^{(\pm)}[\z;\tau_0,t_1]
    =\int_{-\frac{\pi}{2\w}}^{\frac{\pi}{2\w}}dt
    \int_{\e_\tau}^{\infty} d\tau
    \J_>^{(\pm)}(\tau,t)\z'(\tau+\tau_0+\frac{\s}{2},t+t_1).
 \label{si2r}\eq
We now Taylor expand $\z'$ in (\ref{si2l}) and (\ref{si2r}) around
the spacetime points $(\tau_0-\frac{\s}{2},t_1)$ and
$(\tau_0+\frac{\s}{2},t_1)$
respectively; that is, in $S_{<I}^{(\pm)}$ we expand $\z'$ around the
rightmost edge of the continuous region of spacetime to the left of the
low density region, and in $S_{>I}^{(\pm)}$ we expand $\z'$ around
the leftmost edge of the continuous spacetime region to the right of
the low
density region.  Doing this, we derive the following expression,
\bq \z'(\tau+\tau_0\mp\frac{\s}{2},t+t_1)
    =\sum_{m=0}^\infty\sum_{n=0}^\infty
    \frac{1}{m!n!}\tau^mt^n
    \der_{\tau_0}^m\der_{t_1}^n
    \z'(\tau_0\mp\frac{\s}{2},t_1).
 \eq
Substituting this into
(\ref{si2l}) and (\ref{si2r}),
we immediately find
\brr S_{<I}^{(\pm)}[\z;\tau_0,t_1] &=& \sum_{mn}\frac{1}{\w^{m+n+1}}
    h_{<mn}^{(\pm)}
    \der_{\tau_0}^m\der_{t_1}^n\z'(\tau_0-\frac{\s}{2},t_1), \nonumber
\\
     S_{>I}^{(\pm)}[\z;\tau_0,t_1] &=& \sum_{mn}\frac{1}{\w^{m+n+1}}
     h_{>mn}^{(\pm)}
    \der_{\tau_0}^m\der_{t_1}^n\z'(\tau_0+\frac{\s}{2},t_1),
 \err
where
\bq h_{<mn}^{(\pm)}=\frac{\w^{m+n+1}}{m!n!}
    \int_{-\frac{\pi}{2}}^{\frac{\pi}{2}}dt
    \int_{-\infty}^{-\e_\tau} d\tau
    \J_<^{(\pm)}(\tau,t)\tau^mt^n.
 \label{hleft}\eq
and
\bq h_{>mn}^{(\pm)}=\frac{\w^{m+n+1}}{m!n!}
    \int_{-\frac{\pi}{2\w}}^{\frac{\pi}{2\w}}dt
    \int_{\e_\tau}^{\infty}d\tau
    \J_>^{(\pm)}(\tau,t)\tau^mt^n.
 \label{hright}\eq
As we show explicitly in Appendix B, these are computable, finite,
dimensionless numbers.
Furthermore, due to the symmetry properties of $\J_<^{(\pm)}$ and
$\J_>^{(\pm)}$, it follows that
\bq h_{<mn}^{(+)}=(-)^{m+n}h_{>mn}^{(+)}=
    h_{>mn}^{(-)}=(-)^{m+n}h_{<mn}^{(-)}
    \equiv h_{mn}.
 \label{hhh}\eq
Thus,
\brr S_{<I}^{(+)}[\z;\tau_0,t_1] &=&
\frac{1}{\w}h_{00}\z'(\tau_0-\frac{\s}{2},t_1)
    +\frac{1}{\w^2}h_{01}\z''(\tau_0-\frac{\s}{2},t_1)
    +\frac{1}{\w^2}h_{10}\dot{\z}'(\tau_0-\frac{\s}{2},t_1)+\cdots,
    \nonumber\\
    S_{>I}^{(+)}[\z;\tau_0,t_1] &=&
    \frac{1}{\w}h_{00}\z'(\tau_0+\frac{\s}{2},t_1)
    -\frac{1}{\w^2}h_{01}\z''(\tau_0+\frac{\s}{2},t_1)
-\frac{1}{\w^2}h_{10}\dot{\z}'(\tau_0+\frac{\s}{2},t_1)+\cdots,
\nonumber \\
\err
and
\brr S_{<I}^{(-)}[\z;\tau_0,t_1] &=&
\frac{1}{\w}h_{00}\z'(\tau_0-\frac{\s}{2},t_1)
    -\frac{1}{\w^2}h_{01}\z''(\tau_0-\frac{\s}{2},t_1)
-\frac{1}{\w^2}h_{10}\dot{\z}'(\tau_0-\frac{\s}{2},t_1)+\cdots,
\nonumber  \\
   S_{>I}^{(-)}[\z;\tau_0,t_1] &=&
\frac{1}{\w}h_{00}\z'(\tau_0+\frac{\s}{2},t_1)
    +\frac{1}{\w^2}h_{01}\z''(\tau_0+\frac{\s}{2},t_1)
+\frac{1}{\w^2}h_{10}\dot{\z}'(\tau_0+\frac{\s}{2},t_1)+\cdots,
\nonumber  \\
 \err
It is convenient to adopt the following notation,
\bq \z_\pm\equiv\z(\tau_0\pm\frac{\s}{2},t_1).
\label{not}\eq
Since $S_I^{(\pm)}=S_{<I}^{(\pm)}+S_{>I}^{(\pm)}$, it then follows that
\brr S_I^{(+)} &=& \frac{1}{\w}h_{00}(\z_-^{'}+\z_+^{'})
              +\frac{1}{\w^2}h_{01}(\z_-^{''}-\z_+^{''})
              +\frac{1}{\w^2}h_{10}(\dot{\z}_-^{'}-\dot{\z}_+^{'})
              +\frac{1}{\w^3}h_{11}(\dot{\z}_-^{''}+\dot{\z}_+^{''})
              +\cdots \nonumber \\
     S_I^{(-)} &=& \frac{1}{\w}h_{00}(\z_-^{'}+\z_+^{'})
              -\frac{1}{\w^2}h_{01}(\z_-^{''}-\z_+^{''})
              -\frac{1}{\w^2}h_{10}(\dot{\z}_-^{'}-\dot{\z}_+^{'})
              +\frac{1}{\w^3}h_{11}(\dot{\z}_-^{''}+\dot{\z}_+^{''})
              +\cdots.\nonumber \\
 \label{sinl}\err
Using equations (\ref{hleft}), (\ref{hright}), and (\ref{hhh}),
it is straightforward to compute the coefficients $h_{mn}$ and we do it
explicitly  in Appendix B.
Note that due to the cutoff $\e_\tau$ in (\ref{hleft}) and
(\ref{hright}),
these coefficients depend, in general, on $g$.  We find, for instance,
to leading order in $g$, that
\brr h_{00} &=& -\frac{4\sqrt{2}}{9} \nonumber \\
     h_{10} &=& -(\frac{8\pi g}{9})^{1/3} \nonumber \\
     h_{01} &=& -\frac{\pi\sqrt{2}}{9}.
 \label{hee}\err
In general, the $h_{mn}$ are found to have the following $g$
dependence,
\bq h_{mn}\sim\left\{
    \begin{array}{ll}
    g^{m/3} & ;\hspace{.2in} m\le 3 \\
    g & ;\hspace{.2in} m>3 \end{array}\right.
 \label{gee}\eq
Note, from (\ref{sinl}) and (\ref{gee}),
that, as the first index of $h_{mn}$ increases, that the corresponding
terms in $S_I^{(\pm)}$ depend on higher powers of
$g$. However, none of $h_{0n}$ have $g$ dependence for any value of
$n$.
We proceed to analyze the relative impact of these terms on generic
$N$-point
functions.  By putting (\ref{sinl}) back into (\ref{zelta})
we can find all relevant interaction vertices.  These are obtained by
Taylor expanding the exponentials in (\ref{zelta}).  For instance, we
obtain
the quadratic vertices $\frac{1}{\w^2}h_{00}^2\z_-^{'}\z_-^{'}$
and $\frac{1}{\w^3}h_{00}h_{10}\z_-^{'}\z_-^{''}$ where, as discussed
above, $h_{00}\sim 1$ and $h_{10}\sim g^{1/3}$.
It is clear that the effect of the second vertex, containing
$h_{00}h_{10}$,
on any $N$-point function, is suppressed by a factor
$g^{1/3} p/w$, where $p$ is a characteristic momentum,
when compared with effects arising soley from the
first vertex containing $h_{00}^2$.
This is true at tree level. At the quantum level,
 there may be some subtleties to this argument which
we will not discuss in this paper.  Similar considerations
apply to all other induced operators, involving higher $h_{mn}$.
It can thus be shown, provided
\bq p\lapp\w,
 \label{pcon}\eq
that, when working to leading order in $g$, we can
consistently drop all but the $h_{0n}$ terms in (\ref{sinl}).
Now, of the terms which remain, as $n$ increases, the corresponding
terms in $S_I^{(\pm)}$ depend on higher derivatives of $\z$.
Thus, the effect of any vertex, containing $h_{0n}$, on any
$N$-point function, is suppressed by a factor $(p/\w)^n$, relative to
effects arising from vertices containing only $h_{00}$.
If we further restrict momenta, such that
\bq p<<\w,
 \label{rees}\eq
we can then consistently neglect all but the $h_{00}$ terms
in (\ref{sinl}).  This results in a vast simplification of the
final result, so we will assume this approximation.  It would be
completely straightforward, however, to lift the restriction
(\ref{rees}),
and only require (\ref{pcon}).  One would then have to keep all
$h_{0n}$ terms in (\ref{sinl}).
It follows, from (\ref{sinl}), to the order of approximation
given in (\ref{rees}), that
$S_I^{(+)}=S_I^{(-)}$, and therefore that (\ref{zelta}) collapses
to a single exponential.  Plugging (\ref{sinl}) into (\ref{zelta})
and using (\ref{not}) and (\ref{hee}), we then find, to leading order
in $g$,
\bq \D S[\z]=2\w g^{-1/6}e^{-\frac{\pi}{2 g}}\int dt_1
    \exp\bl\frac{4\sqrt{2}}{3\w}\bpl\z'(\tau_0+\frac{\s}{2},t_1)
    +\z'(\tau_0-\frac{\s}{2},t_1)\bpr\br.
 \label{nl}\eq
Note however that equation (\ref{nl}) includes nonlocal interactions,
since it involves contributions coming from
$\z'$ evaluated simultaneously at $\tau_0-\frac{\s}{2}$ and also at
$\tau_0+\frac{\s}{2}$.  This is not suprising though, since we
have arrived at this result by integrating over single eigenvalue
instantons, which link effects on the left-hand side of the low-density
region with effects on the the right-hand side of this region, and
because there
is a finite seperation between these two sectors. One may wish
to find some further approximation which would render the effective
theory
local.  This can be done as follows.  Provided we
consider momenta which satisfy (\ref{rees}), and provided also that
$\w\lapp\frac{1}{\s}$, the effective width of the low density region
as seen by any field will be essentially zero.  We therefore Taylor
expand $\z'(\tau_0\pm\frac{\s}{2},t_1)$ around the point
$(\tau_0,t_1)$,
thereby
taking
\bq \frac{1}{\w}\z'(\tau_0\pm\frac{\s}{2},t_1)
    =\frac{1}{\w}\z'(\tau_0,t_1)\pm\frac{\s\w}{2\w}\z''(\tau_0,t_1)
    +\cdots.
 \eq
Then, in a manner identical to the previous discussion,
we find that the
contributions coming from vertices which involve $\s$ are always
suppressed
by $(\s\w)p/\w$, where $p$ is a characteristic momentum.
Note that, since we now assume
$\w\lapp\frac{1}{\s}$, the factor $(\s\w)$ is $\lapp\O(1)$.
So, provided that
\bq p<<\w\lapp\frac{1}{\s},
 \eq
we may write the lowest order instanton-induced change
in the collective field action approximately, in local form, as
follows,
\bq \D S[\z]=2\w g^{-1/6}e^{-\frac{\pi}{2g}}\int dt
     e^{-\frac{2\sqrt{2}}{3\w}\z'(\tau_0,t)}.
 \eq
We have dropped the subscript $``1"$ on $t_1$ because it is now
superfluous.  This result can be written as a two-dimensional
integral over a density $\D S=\int dt d\tau \D\L$, where
\bq \D\L=2\w g^{-1/6}e^{-\frac{\pi}{2g}}
    \delta(\tau-\tau_0)e^{-\frac{2\sqrt{2}}{3\w}\z'(\tau,t)}.
 \eq
This is the final result of our calculation.

\renewcommand{\theequation}{4.\arabic{equation}}
\setcounter{equation}{0}
{\fl{\bf 4. Conclusion}}\\
\indent
We have presented a detailed analysis of the interplay between the
continuous
and discrete sectors of a $d=1$ bosonic matrix model, and have
performed an
explicit and complete calculation of the single eigenvalue instantons
in the theory.  In addition we have derived the precise form of the
lowest
order operators which are induced in the theory when the instantons are
integrated out.
The relevant fact which we have demonstrated
is that the nonperturbative aspects of the collective field theory
can be isolated, and their leading order effects systematically
incorporated.
This calculation is an essential preliminary for an
interesting analysis involving  the $d=1, {\cal N}=2$ supersymmetric
matrix model.  The supersymmetric case will be presented in a
forthcoming
paper\cite{nbfo}.

\renewcommand{\theequation}{A.\arabic{equation}}
\setcounter{equation}{0}
{\fl{\bf Appendix A: Calculation of $\M$}}\\
\indent
In this Appendix we compute $\M$, the mass scale characteristic of
nonperturbative effects in the collective field theory.  In section 3
we
found that
\bq \M=\sqrt{\frac{\pi}{2 g}}\sqrt{\frac{\det \O_0}{{\det}'\O_1}}
    e^{-\frac{\pi}{2g}},
 \label{mmm}\eq
which was first stated as equation (\ref{mdef}).  The first factor in
this
expression is a functional jacobian which results from the extraction
of the
zero mode $t_0$ from the $[d\l_0]$ functional measure, and the last
term
is a fugacity factor.  The middle term includes the quantum effects
involving $\tilde{\l}$, the fluctuations of $\l_0$ around the instanton
background.  It is the result of performing the $\int [d\tilde{\l}]$
integration.  The operators $\O_0$ and $\O_1$ are given in
(\ref{oodef})
and (\ref{oidef}) respectively.
The first operator is given by
\bq \O_0 = -\der_t^2-\w^2+
     \int dx\frac{3\tilde{\vphi}_0^{'}(x)}{(x-\l_{\so}^{(+)})^4}.
 \label{oo}\eq
But
\brr \int dx\frac{3\tilde{\vphi}_0^{'}(x)}{(x-\l_{\so})^4}
     &=& \frac{6\w}{\pi}\int_{A+\e_x}^\infty
     dx\frac{\sqrt{x^2-A^2}}{(x-A)^4} \nonumber \\
     &=& \frac{6\w}{\pi}\frac{1}{15 A^2}\bl 1+6\sqrt{2}(\frac{A}{\e_x})
     ^{\frac{5}{2}}(1+\O(\frac{\e_x}{A}))\br \nonumber \\
     &\approx & \frac{32}{5(3\pi^4)^{2/3}}\frac{\w^2}{g^{2/3}}.
 \label{oop}\err
The approximation made in the last line of (\ref{oop}) is valid
provided $g<<1$, which we always assume.
Therefore we can reexpress $\O_0$ as follows,
\bq \O_0=-\der_t^2+\a\w^2,
 \eq
where
\bq \a= \frac{32}{5(3\pi^4g)^{2/3}}-1.
 \label{adef}\eq
The second operator is given by
\bq \O_1 = -\der_t^2-\w^2+
     \int dx\frac{3\tilde{\vphi}_0^{'}(x)}{(x-\l_{{0}}^{(+)})^4}.
 \label{oi}\eq
where $\l_{{0}}^{(+)}=A\sin\w(t-t_1)$.  For convenience, for this
particular
calculation, we define $\g=\sin\w(t-t_1)$.  Thus,
\brr \int dx\frac{3\tilde{\vphi}_0^{'}(x)}{(x-\l_{{0}}^{(+)})^4}
     &=& \int dx\frac{3\tilde{\vphi}_0^{'}(x)}{(x-\g A)^4} \nonumber \\
     &=& \frac{3\w}{\pi A^2}
     \frac{\g}{(1-\g^2)^{5/2}}\bl
     \bpl 1+\g \bpr +
     \bpl\tan^{-1}\frac{\g}{\sqrt{1-\g^2}}-\frac{\pi}{2}\bpr \br
\nonumber \\
     &\approx& \frac{\g}{(1-\g^2)^{5/2}} \cdot \w^2 g \nonumber \\
     &<<& \w^2.
 \err
The last line is true provided $\g$ isn't too close to one, which we
assume
since this operator only applies when acting on a discrete
eigenvalue.
So, for a crude but reasonable calculation, we can neglect
the last term in (\ref{oi}) relative to the pure $\w^2$ term.
Therefore, we take
\bq \O_0=-\der_t^2-\w^2.
 \eq
We find the spectrum associated with both $\O_0$ and $\O_1$ by
solving the eigenvalue problem,
\bq \O_i\l_n=\w_n^{(i)2}\l_n,
 \eq
where $i=0$ or $1$ and the $\l_n(t)$ are defined over
$-\frac{\pi}{2\w}\le t\le\frac{\pi}{2\w}$,
satisfy the boundary condition
$\l_n(t=\pm\frac{\pi}{2\w})=0$, and are orthonormal,
\bq \int_{-\frac{\pi}{2\w}}^{-\frac{\pi}{2\w}}dt\l_n(t)\l_m(t)=\d_{nm}.
 \eq
Both operators have the same set of eigenfunctions, which are
\bq \l_n(t)=\sqrt{\frac{2\w}{\pi}}\sin\bl n\w(t-\frac{\pi}{2\w})\br.
 \eq
It is easily seen that
\bq \w_n^{(0)2}=(n^2+\a)\w^2
 \eq
and
\bq \w_n^{(1)2}=(n^2-1)\w^2.
 \eq
Thus,
\brr \det \O_0 &=& \prod_{n=1}^\infty (n^2+\a)\w^2 \nonumber \\
     {\det}' \O_1 &=& \prod_{n=2}^\infty (n^2-1)\w^2.
 \err
In the $\det'$ case, we have removed the zero eigenvalue $\w_1^{(1)2}$.
This gives the following result,
\bq \sqrt{\frac{\det\O_0}{\det'\O_1}}=\w\sqrt{(1+\a)\prod_{n=2}^\infty
    (\frac{n^2+\a}{n^2-1})},
 \label{mere}\eq
where $\a$ is given in (\ref{adef}).

Now, in order that we respect assumptions made in section 2,
specifically equation (\ref{gdef}), we must take $g<<1$.  However,
since the factor $\exp(-\frac{\pi}{2g})$ which appears in
(\ref{mmm}) rapidly becomes incredibly small as $g$ becomes
smaller than $.01$, where it has a value $\sim 10^{-68}$, we consider
a ``reasonable" range of $g$ to be between $.01$ and $.1$.
In this way we consider circumstances in line with our assumptions but
which
don't allow such a supression of instanton effects as to make them
physically uninteresting.  We point out that for $g=.1$ and $.05$,
$\exp(-\frac{\pi}{2g})$ is $\sim 10^{-7}$ and $\sim 10^{-14}$
respectively.  Now, we have evaluated (\ref{mere}) numerically
for various reasonable values of $g$, and we find that for $g=.1$,
$.05$,
and $.01$, that equation (\ref{mere}) becomes $1.03\w$, $1.49\w$, and
$4.56\w$ respectively.  Since these values are all $\w$ times a factor
of $\O(1)$, and since it is difficult to obtain a more compact
closed-form expression for equation (\ref{mere}) which is valid
over the ``reasonable" range of $g$, it is useful, over this
range of $g$, to simply take
\bq \sqrt{\frac{\det\O_0}{\det'\O_1}}\approx\w,
 \eq
which we will do for definiteness.
Using (\ref{mmm}), we then arrive at the following result
\bq \M\approx {\w}\sqrt{\frac{\pi}{2 g}}e^{-\frac{\pi}{2g}}.
 \label{scale}\eq
To conclude,
in this Appendix we have shown, regardless of these concerns, that for
small
values of $g$, the characteristic nonperturbative mass scale in
the collective field theory is as given in equation (\ref{scale}).

\renewcommand{\theequation}{B.\arabic{equation}}
\setcounter{equation}{0}
{\fl{\bf Appendix B: Calculation of $h_{mn}$}}\\
\indent
In this appendix we calculate the leading order behaviour
of the coefficients $h_{mn}$.  From Eq.(\ref{hhh}) we see that it is
enough to
calculate
\bq h_{>mn}^{(+)}=\frac{\w^{m+n+1}}{m!n!}
    \int_{-\frac{\pi}{2\w}}^{\frac{\pi}{2\w}}dt
    \int_{\e_\tau}^{\infty}d\tau
    \J_>^{(+)}(\tau,t)\tau^mt^n.
 \label{ba}\eq
First, rescale $q=\w t$, that simply takes away $n+1$ powers of $\w$
and sets
the integration boundaries to $-\pi/2, +\pi/2$ then substitute the
expression
for $\J_>^{(+)}$ from Eq.(\ref{si2r}) into Eq.(\ref{ba})
\bq h_{mn}=\frac{\w^{m}}{m!n!}
    \int_{-\frac{\pi}{2}}^{\frac{\pi}{2}}dq
    \int_{\e_\tau}^{\infty}d\tau[
    \frac{1}{(x(\tau-\tau_0-\frac{\s}{2})-\l_0^{(+)}(q))^2}-
    \frac{1}{(x(\tau-\tau_0-\frac{\s}{2})-\l_{\emptyset}^{(+)}(q))^2}]\
\tau^m
q^n.
 \label{bb}\eq
Now,  substitute the expression for $x(\tau-\tau_0-\frac{\s}{2})$ and
rescale
$r=\w\tau$ to obtain
\bq h_{mn}=\frac{1}{m!n!}\frac{1}{\w}
    \int_{-\frac{\pi}{2}}^{\frac{\pi}{2}}dq
    \int_{\e_r}^{\infty}dr[
    \frac{1}{(A \cosh\ r-\l_0^{(+)}(q))^2}-
    \frac{1}{(A \cosh\ r-\l_{\emptyset}^{(+)}(q))^2}]\ r^m q^n,
 \label{bc}\eq
where,  using Eq.(3.62),  $\e_r=\frac{1}{\sqrt{2}}(3\pi g)^{1/3}$.
Note that $\e_r$ is a dimensionless number.
Then substitute the expressions for $\l_0^{(+)}(q)$,
$\l_{\emptyset}^{(+)}(q)$
to obtain
\brr h_{mn}&=&\frac{g}{m!n!}\bbl
    \int_{-\frac{\pi}{2}}^{\frac{\pi}{2}}dq
    \int_{\e_r}^{\infty}dr
    \frac{1}{(\cosh\ r- \sin (q))^2}\ r^m q^n\nonumber \\
&-&\int_{-\frac{\pi}{2}}^{0} dq
    \int_{\e_r}^{\infty}dr
\frac{1}{(\cosh\ r+1)^2}\ r^m q^n
\nonumber \\ &-&\int_{0}^{\frac{\pi}{2}} dq
    \int_{\e_r}^{\infty}dr
    \frac{1}{(\cosh\ r-1)^2}\ r^m q^n\bbr.
 \label{bd}\err
where we used $g=\frac{1}{\w A^2}$. From Eq.(\ref{bd}) we can see that
$h_{mn}$
are finite dimensionless numbers for all $m,n$.

To estimate the leading $g$ dependence in the small $g$ limit to the
coefficients $h_{mn}$ note that the main contribution to $h_{mn}$, for
$m\le
3$,  comes
from
regions that are close to the lower boundary of integration. In this
region,
the third term in the previous equation is always much larger than the
first
two terms.
We can then write, to leading order in $g$,
\brr
h_{mn}&=&-\frac{g}{m!n!}\int_{0}^{\frac{\pi}{2}} dq
    \int_{\e_r}^{\infty}dr
    \frac{1}{(\cosh\ r-1)^2}\ r^m q^n \nonumber \\
&=& -\frac{g (\pi/2)^{n+1}}{m!(n+1)!}
    \int_{\e_r}^{\infty}dr
    \frac{1}{(\cosh\ r-1)^2}\ r^m.
\err
The leading $g$ dependence of the coefficients, for $m\le 3$, can
therefore be
calculated to
be
\bq h_{mn}\approx  g\ \e_r^{m-3}=g^{m/3}.\eq
For $m>3$, all the terms in Eq.(\ref{bd}) are equally important and, to
leading
order, their $g$-dependence is give by the overall factor of $g$. The
actual results for the first few
coefficients are give by
\brr h_{00}&=&-\frac{4\sqrt{2}}{9}\nonumber \\
h_{10}&=&-(\frac{8 \pi g}{9})^{1/3}\nonumber \\
h_{01} &=&-\frac{\pi\sqrt{2}}{9}
\err

\end{document}